\documentclass{aa}
\usepackage{txfonts}
\usepackage{graphicx}
\usepackage{subcaption}
\usepackage{threeparttable} 
\usepackage[colorlinks=true,allcolors=blue]{hyperref}
\usepackage{gensymb}

\DeclareUnicodeCharacter{5415}{}  
\DeclareUnicodeCharacter{5EFA}{}  
\DeclareUnicodeCharacter{4F1F}{}  

\begin{document} 

\authorrunning{Mountrichas et al.}
\titlerunning{AGN incidence and morphology in DESI–eRASS1}

\title{The incidence and star-formation properties of X-ray AGN across the star-forming main sequence in DESI–eRASS1: the role of host-galaxy morphology}

\author{G. Mountrichas\inst{1}, F. J. Carrera\inst{1}, V. A. Masoura\inst{1}, S. Mateos\inst{1}, M. Siudek\inst{2, 3}, A. Corral\inst{1} }
          
     \institute {Instituto de Fisica de Cantabria (CSIC-Universidad de Cantabria), Avenida de los Castros, 39005 Santander, Spain.
              \email{gmountrichas@gmail.com}
              \and
       Instituto de Astrof\'{\i}sica de Canarias, V\'{\i}a L\'actea, 38205 La Laguna, Tenerife, Spain\label{aff1}
\and
Instituto de Astrof\'isica de Canarias (IAC); Departamento de Astrof\'isica, Universidad de La Laguna (ULL), 38200, La Laguna, Tenerife, Spain\label{aff2}}

\abstract{\noindent
\textit{Context.} The interplay between star formation and supermassive black-hole growth is central to galaxy evolution. While X-ray selected AGN preferentially reside in star-forming galaxies, the role of host-galaxy morphology in regulating both star-formation enhancement and AGN triggering across the star-forming main sequence remains unclear.

\noindent
\textit{Aims.} We investigate the star-formation properties and incidence of X-ray AGN across the star-forming main sequence using the DESI–eRASS1 dataset, focusing on the role of host-galaxy structure.

\noindent
\textit{Methods.} Our analysis includes 1\,171 X-ray selected AGN and 45\,374 non-AGN star-forming galaxies at $z \leq 1.5$. We quantify star formation in AGN hosts relative to matched control samples using the $\mathrm{SFR}_{\rm norm}$ parameter and examine its dependence on X-ray luminosity ($L_{\rm X}$) and specific accretion rate ($\lambda_{\rm sBHAR}$). We also measure AGN incidence as a function of distance from the star-forming main sequence ($\Delta{\rm MS}$), separating disk- and spheroid-dominated systems based on Sérsic-index classifications.

\noindent
\textit{Results.} $\mathrm{SFR}_{\rm norm}$ remains close to unity at low to intermediate $L_{\rm X}$ and increases at higher luminosities, with the transition shifting toward higher $L_{\rm X}$ in more massive systems. This trend depends on morphology: in the stellar-mass range $10.5 \le \log(M_\star/{\rm M_\odot}) < 11.5$, disk-dominated AGN hosts exhibit enhanced $\mathrm{SFR}_{\rm norm}$ at moderate $L_{\rm X}$, while spheroid-dominated systems remain consistent with unity. No comparable morphology dependence is found when using $\lambda_{\rm sBHAR}$. The incidence of X-ray AGN increases strongly with $\Delta{\rm MS}$ in both redshift bins, with a steeper dependence at higher redshift. The $\Delta{\rm MS}$–incidence relation is also morphology-dependent and evolves with redshift.

\noindent
\textit{Conclusions.} The connection between star formation and AGN activity is primarily governed by global gas availability but modulated by host-galaxy structure and cosmic epoch. Absolute AGN power is more tightly linked to host-wide star formation than accretion efficiency normalised by stellar mass.}

\keywords{}
   
\maketitle  

\section{Introduction}

Understanding the connection between star formation and supermassive black-hole (SMBH) growth remains one of the central challenges in galaxy evolution studies. Over the past two decades, large multi-wavelength surveys have established that most massive galaxies host SMBHs at their centres, and that the growth of these black holes is closely linked to the evolution of their host galaxies (e.g. \citealt{Kormendy2013}; \citealt{Heckman2014}). Scaling relations between SMBH mass and host-galaxy properties, such as stellar mass, $M_\star$, and bulge velocity dispersion, suggest a long-term co-evolution between galaxies and their central engines. However, the physical mechanisms that couple star formation and black-hole accretion, and the timescales over which this coupling operates, remain actively debated.

One powerful approach to probing this connection is to compare the star-formation properties of active galactic nuclei (AGN) hosts to those of inactive galaxies matched in $M_\star$ and redshift. Several earlier studies reported that X-ray selected AGN
preferentially occupy the green valley or lie below the star-forming
main sequence, suggesting a possible connection between AGN activity
and the transition from star-forming to quiescent systems
(e.g. \citealt{Nandra2007}; \citealt{Treister2009}; \citealt{Povic2012};
\citealt{Mahoro2017}). More recent studies, however, have shown that X-ray selected AGN typically reside in star-forming galaxies and broadly trace the star-forming main sequence (e.g. \citealt{Mullaney2015}; \citealt{Aird2018, Aird2019}). Nevertheless, whether AGN hosts exhibit enhanced, suppressed, or typical star formation relative to non-AGN galaxies at fixed host properties depends on the AGN luminosity and $M_\star$  considered (e.g. \citealt{Mountrichas2021c, Mountrichas2022a, Mountrichas2022b, Mountrichas2024a}; \citealt{Cristello2024}). 

A complementary perspective is provided by the incidence of AGN activity across the star-forming main sequence. Rather than asking whether AGN hosts have elevated star formation, this approach asks how the probability of hosting an AGN depends on a galaxy’s position relative to the main sequence. Recent work has demonstrated that AGN incidence increases strongly with star-formation rate (SFR) at fixed M$_\star$, suggesting that black-hole accretion is closely linked to the availability of cold gas (e.g. \citealt{Aird2018, Aird2019}). However, the role of host-galaxy structure in regulating this connection remains less well constrained.

Morphology provides critical information about the dynamical state and gas distribution of galaxies. Disk-dominated systems are typically gas-rich and rotationally supported, favouring sustained, secular star formation, while spheroid-dominated galaxies are generally more centrally concentrated and dynamically hotter, conditions that may regulate the efficiency with which gas is transported toward the nucleus. Observational studies have reported structural differences between AGN hosts and inactive galaxies, as well as between bulge- and disk-dominated AGN systems (e.g. \citealt{Schawinski2010, Cisternas2011, Kocevski2012, Simmons2013, Bruce2014, ZouYang2019, Ni2021}). While moderate-luminosity AGN are frequently found in disk galaxies, particularly at intermediate redshift, evidence for morphology-dependent trends in star-formation activity and black-hole growth remains mixed. More recently, studies have begun to explore differences in stellar populations and structural properties between bulge- and disk-dominated AGN hosts (e.g. \citealt{Mountrichas2022b}). However, a systematic investigation of how morphology modulates both the star-formation enhancement of AGN hosts and the incidence of AGN across the star-forming main sequence remains limited by sample size, homogeneous multi-wavelength coverage, and well-characterised X-ray selection.

The combination of the \textit{eROSITA} all-sky X-ray survey with wide-area optical surveys provides an unprecedented opportunity to revisit these questions with large and homogeneous samples. 
In this work we combine X-ray sources from the first \textit{eROSITA} all-sky survey data release (eRASS1; \citealt{Merloni2024}) with galaxy catalogues from the Dark Energy Spectroscopic Instrument (DESI) survey and the Legacy Imaging Surveys. 
The Legacy Surveys provide deep optical photometry and structural measurements, including Sérsic indices that allow morphological classification of galaxies \citep{Dey2019}. 
Spectroscopic redshifts are obtained from the DESI survey \citep{DESI2022}, enabling accurate distance measurements over wide cosmological volumes. 
Together, these datasets enable consistent measurements of $M_\star$, SFR, morphology, and AGN luminosity across a wide redshift range.

In this work, we investigate both the star-formation properties and the incidence of X-ray AGN across the star-forming main sequence using the DESI–eRASS1 dataset. We quantify star formation in AGN hosts relative to matched non-AGN control samples through the normalized SFR, $\mathrm{SFR}_{\rm norm}$, parameter, and we examine how this quantity depends on X-ray luminosity ($L_{\rm X}$) and specific accretion rate ($\lambda_{\rm sBHAR}$). In parallel, we measure the probability that a star-forming galaxy hosts an X-ray AGN as a function of its offset from the main sequence ($\Delta {\rm MS}$). In both cases, we explicitly explore the role of host-galaxy morphology by separating disk- and spheroid-dominated systems.

This dual approach allows us to address three key questions: (i) does enhanced star formation in AGN hosts depend on galaxy structure? (ii) is AGN triggering more closely linked to absolute AGN power or to accretion efficiency? and (iii) how does the interplay between star formation, morphology, and AGN activity evolve with redshift? By combining star-formation normalisation and AGN-incidence analyses within a single, homogeneous framework, we aim to provide a unified picture of the physical processes that regulate SMBH growth in different structural and evolutionary contexts.

The structure of this paper is as follows. In Sect. \ref{sec:data} we describe the DESI and eRASS1 datasets. Section \ref{sec:analysis} details the sample selection, SED fitting, morphological classification, and completeness corrections. The main results are presented in Sect.~\ref{sec:results}. 
In Sect.~\ref{sec:discussion} we discuss the physical implications of our findings in the context of previous work. Our main findings are summarised in Sect. \ref{sec:summary}.

Throughout this work, we assume a flat $\Lambda$CDM cosmology with
$H_0 = 70\,{\rm km\,s^{-1}\,Mpc^{-1}}$, $\Omega_{\rm M}=0.3$, and
$\Omega_\Lambda=0.7$.

\section{Data}
\label{sec:data}

This work is based on the combination of optical spectroscopic and photometric data from the Dark Energy Spectroscopic Instrument (DESI) survey with X-ray data from the first all-sky survey performed by \textit{eROSITA} (eRASS1). Below we briefly describe the two datasets and the value-added catalogues used throughout this paper.

\subsection{The DESI survey}
\label{sec:desi}

DESI is a wide-area spectroscopic survey conducted on the 4-m Mayall telescope at Kitt Peak, designed to obtain optical spectra for tens of millions of galaxies and quasars over $\sim 14\,000~\mathrm{deg}^2$ of the extragalactic sky \citep{DESICollaboration2016, DESICollaboration2023}. In addition to redshift measurements, DESI provides extensive multi-wavelength photometry compiled from several imaging surveys, enabling detailed studies of galaxy physical properties across cosmic time.

In this work we make use of the DESI value-added catalogue (VAC) presented by \citet{Siudek2024}, which provides photometry and homogeneous estimates of $M_\star$, SFR, and other galaxy properties derived through SED fitting using \textsc{CIGALE} \citep[][]{Boquien2019}. The catalogue is based on DESI Data Release 9 (DR9) imaging and includes extensive quality flags and ancillary information that allow the construction of well-defined galaxy samples.

We do not use the $M_\star$, SFR or AGN
fractions provided by the VAC. Instead, these quantities are derived
independently in this work through SED fitting with \textsc{CIGALE},
using the same templates, parameter grid, and methodology adopted in
our previous studies (see Sect. \ref{sec:sed}). The VAC is therefore used as the photometric and ancillary input catalogue.

The parent DESI VAC contains approximately $1.34 \times 10^6$ sources with reliable photometry and redshift information. We apply a series of photometric quality cuts, signal-to-noise requirements, and redshift constraints to define a clean galaxy sample suitable for SED fitting and for cross-matching with the \textit{eROSITA} data. These selection criteria are described in detail in Sect.~\ref{sec:selection}.


\subsection{The eRASS1 X-ray AGN sample}
\label{sec:erass}

The X-ray sample used in this study is drawn from the first
all-sky survey conducted by eROSITA onboard the Spectrum-Roentgen-Gamma
mission (eRASS1; \citealt{Merloni2024}). The eRASS1 Main catalogue
contains 930\,203 X-ray sources detected in the 0.2--2.3 keV band
over the Western Galactic Hemisphere. Following \citet{Salvato2025},
we restrict the parent X-ray sample to point-like eROSITA detections
(\texttt{EXT\_LIKE}=0), which correspond to 903\,521 sources \citep[for more details see Sect.~2.1 in][]{Salvato2025}.

We use the counterpart-identification catalogues presented by
\citet{Salvato2025}, which associate eRASS1 sources with optical and
infrared counterparts using Legacy Survey DR10 (LS10), Gaia DR3, and
CatWISE2020 information. This catalogue
provides optical counterparts within the DESI/Legacy Survey footprint
and is therefore the most appropriate product for cross-matching with
the DESI VAC. The cross-match with the DESI VAC is performed using the
optical counterpart positions provided by \citet{Salvato2025}.

X-ray fluxes are provided in several energy bands; throughout this work we make use of the soft-band (0.2-2.3\,keV) measurements and convert them to rest-frame $2$--$10~\mathrm{keV}$ luminosities  assuming a power-law spectrum with photon index $\Gamma = 1.8$. K-corrections are applied consistently to all sources. Specifically, we compute

\begin{equation}
L_{\rm X}(2-10\,{\rm keV}) =
4\pi d_{\rm L}^{2} \, f_{0.2-2.3}
(1+z)^{\Gamma-2}
\frac{\int_{2}^{10} E^{1-\Gamma}\,dE}
{\int_{0.2}^{2.3} E^{1-\Gamma}\,dE},
\end{equation}
where $d_{\rm L}$ is the luminosity distance and $f_{0.2-2.3}$ is the
observed eROSITA soft-band flux. The final factor accounts for the
conversion between the observed soft band and the rest-frame 2--10 keV
band under the assumed power-law spectral shape.

We note that these luminosities are approximate and do not account for
source-to-source variations in spectral slope or intrinsic absorption.
However, adopting a uniform conversion provides a homogeneous luminosity
estimate for the full eRASS1-DESI sample. This is preferable to mixing
soft- and hard-band luminosity estimates, which would introduce a
heterogeneous selection function because hard-band detections are
available only for a subset of the sources.
 
We define our X-ray AGN sample by imposing a rest-frame
2--10 keV luminosity threshold of
$\log(L_{\rm X}/{\rm erg\,s^{-1}}) > 42$. This threshold is widely
used in X-ray AGN studies to minimise contamination from X-ray
emission associated with stellar processes, hot gas, or X-ray binaries
(e.g. \citealt{Brandt_Alexander2015,Aird2018}). We therefore refer to
these sources as X-ray selected AGN throughout the paper. We note,
however, that the Salvato et al. catalogues provide high-reliability
counterparts to eROSITA X-ray sources rather than optical spectroscopic
AGN classifications; our AGN definition is based on the X-ray luminosity
criterion.

All analyses presented in this paper are restricted to the redshift range $0 \leq z \leq 1.5$, where the DESI photometry and the eRASS1 AGN selection are both well characterised. At higher redshifts, increasing surface-brightness dimming and angular-resolution limitations lead to less reliable morphological classifications, motivating our adopted upper redshift limit.

\section{Analysis}
\label{sec:analysis}

This section describes the methodology adopted to derive the physical properties of galaxies and AGN, define the final analysis samples, and compute the quantities used in the results presented in Sect.~\ref{sec:results}. Particular emphasis is placed on consistency with our previous studies, in order to enable meaningful comparisons.

\subsection{SED fitting and identification of SED-selected AGN}
\label{sec:sed}

We estimate galaxy physical properties by performing our own SED fitting with the \textsc{CIGALE} code \citep{Boquien2019, Yang2020, Yang2022}. The M$_\star$, SFRs, AGN fractions, and AGN bolometric luminosities
used throughout this work are therefore not taken from the Siudek et al.
VAC, but are derived from our CIGALE runs. The adopted SED-fitting setup is identical to that used in our previous works (e.g. \citealt{Mountrichas2021b, Mountrichas2021c, Mountrichas2023a, Mountrichas2023c, Mountrichas2024b, Mountrichas2024c, Mountrichas2024d}), ensuring full consistency in the derived $M_\star$, SFRs, and AGN-related parameters.

In brief, the stellar component of the galaxies is described using a delayed star-formation history (SFH) of the form $\mathrm{SFR}(t) \propto t \times \exp(-t/\tau)$. In addition, a secondary burst component is included, modelled as a constant episode of star formation lasting 50 Myr, following the prescriptions of \citet{Malek2018} and \citet{Buat2019}.

Stellar emission is modelled using the single stellar population (SSP) templates of \citet{Bruzual_Charlot2003} and is attenuated according to the dust attenuation law of \citet{Charlot_Fall_2000}. Nebular emission is incorporated using templates based on \citet{VillaVelez2021}. The infrared emission from dust heated by stars is represented using the models of \citet{Dale2014}, explicitly excluding any AGN contribution.

AGN emission is included through the \textsc{SKIRTOR} models \citep{Stalevski2012, Stalevski2016}, which self-consistently account for both the dusty torus and the central engine. The full set of SED modules and the adopted parameter space are identical to those used in our previous studies and are described in detail in Table~1 of \citet{Mountrichas2022b, Mountrichas2022c}.

Following our standard approach, we identify SED-selected AGN based on the AGN fractional contribution to the infrared luminosity (AGN fraction, $frac_{AGN})$. Sources with a significant AGN contribution \citep[$frac_{AGN}\geq 0.2$, e.g.][]{Mountrichas2021c, Mountrichas2024a}  are flagged as SED AGN and excluded from the non-AGN control sample used in the computation of star-formation normalisation (Sect.~\ref{sec:sfrnorm}). This procedure minimises the inclusion of galaxies hosting radiatively significant AGN in the control sample.

\subsection{Morphological classification}
\label{sec:morphology}

Morphological information is derived from the photometric modelling of DR9 of the Legacy Imaging Surveys \citep{Dey2019}, which provide deep optical imaging over the DESI footprint. Structural parameters are obtained from single-component parametric fits to the galaxy surface-brightness distribution performed with the Tractor image-modelling pipeline \citep{Lang2016}. These measurements are distributed through the DESI value-added catalog products and are accessed here via the \texttt{zall-pix} catalog.

In this work we use the Sérsic index, $n$, as a first-order characterisation of galaxy structure. The Sérsic index has been widely used as a proxy for distinguishing between disk- and bulge-dominated systems in large galaxy surveys, with low values typically associated with disk-dominated galaxies and high values with spheroid-dominated systems.

We classify galaxies into three morphological categories based on their single-component $n$ values:

\begin{itemize}
    \item disk-dominated systems, defined as galaxies with $n < 2$;
    \item spheroid-dominated systems, defined as galaxies with $n \geq 4$;
    \item intermediate systems, corresponding to the transition regime with $2 \leq n < 4$.
\end{itemize}

The structural parameters are matched to our working catalog using the unique DESI identifiers (\texttt{TARGETID}, \texttt{SURVEY}, \texttt{PROGRAM}, and \texttt{HEALPIX}). Intermediate systems are excluded from analyses where a clear disk--spheroid comparison is required, both because they represent a minor fraction of the sample and because their structural properties span a broad transition regime.

Although single-Sérsic fits provide only a simplified description of galaxy structure, they have been shown to provide a robust first-order separation between disk- and spheroid-dominated systems in large imaging surveys. 
In Appendix~\ref{sec:appendix_morphology} we illustrate the distribution of Sérsic indices as a function of M$_\star$ for the galaxies in our sample, demonstrating that the adopted thresholds ($n<2$ for disks and $n\geq4$ for spheroids) effectively isolate distinct structural regimes. The final numbers of disk- and spheroid-dominated galaxies after all SED,
redshift, quiescent-galaxy, and AGN selection criteria are given in
Sect.~\ref{sec:sfrnorm}.

\subsection{Sample selection}
\label{sec:selection}

In this subsection, we describe the various criteria applied to construct the final dataset used in our analysis, ensuring well-defined samples with reliable photometric, structural, and physical properties.

\subsubsection{Photometric quality, redshift, and footprint selection}
\label{sec:photoselection}

Starting from the full DESI VAC, we apply a series of photometric and quality cuts to define a clean parent galaxy sample suitable for SED fitting and for cross-matching with the eRASS1 AGN catalogue.

Specifically, we require:
\begin{itemize}
    \item positive flux measurements in the optical ($g$, $r$, $z$) and mid-infrared (WISE W1, W2, and W4) bands used for SED fitting;
    \item \texttt{FLAGOPTICAL = 3}, indicating high-quality optical photometry;
    \item a signal-to-noise ratio $\geq 3$ in the WISE bands (W1, W2, W4);
    \item a redshift in the range $0 \leq z \leq 1.5$.
\end{itemize}

Applying these criteria sequentially reduces the initial DESI sample of $1,337,250$ sources to $354,836$ galaxies. We then restrict this sample to the eRASS1 footprint, yielding a final parent sample of $138,768$ galaxies, which constitutes the input catalogue for the SED fitting. Although DESI covers a wider area of the sky, limiting the analysis to the eRASS1 footprint ensures that the parent galaxy sample is uniformly exposed to X-ray observations, which is essential for defining unbiased AGN and non-AGN samples.

We note that all 138,768 sources entering the SED fitting have positive
fluxes in the optical $g$, $r$, and $z$ bands from the Legacy Surveys, as
well as in the mid-infrared WISE W1, W2, and W4 bands. The optical photometry is based on
the Legacy Survey Tractor model photometry, while the WISE measurements
correspond to forced photometry at the optical source positions. Therefore,
the photometry is not based on a single fixed aperture, but on the
model/forced-photometry framework of the Legacy Survey catalogues. The median relative photometric uncertainties, estimated as
$\sigma_f/f$ using the inverse-variance columns, are 1.2, 0.8, 0.8, 1.9, 5.3,
and 9.1 per cent in $g$, $r$, $z$, W1, W2, and W4, respectively, for the 138,768 sources used in the SED fitting.

Furthermore, we do not require our sources to have fluxes in the WISE W3, since our previous works \citep[e.g.,][]{Mountrichas2021b, Mountrichas2021c, Mountrichas2022a, Mountrichas2022b, Mountrichas2024a} have shown that the combination of optical photometry with W1, W2,
and W4 provides the constraints required by our adopted templates on the
stellar component and the mid-infrared dust/AGN emission. W3 is therefore
not required for the physical parameters used in this analysis.

Finally, at higher redshifts (e.g. $z>0.5$), the emission from young stellar populations can be constrained effectively by the optical bands, since they probe progressively shorter rest-frame wavelengths, including the near-ultraviolet (UV). At lower redshifts, however, shorter-wavelength observations can provide additional leverage for tracing recent star formation. Nevertheless, \citet{Koutoulidis2022} showed that the absence of UV photometry does not significantly affect the reliability of SFR estimates derived with \textsc{CIGALE}. For this reason and also to preserve consistency with the SED-fitting configuration adopted in our previous AGN host-galaxy studies we do not include UV photometry in the SED fitting process.

\subsubsection{SED-fitting and morphology quality criteria}
\label{sec:sedmorphcuts}

To ensure robust estimates of host-galaxy physical properties, we apply a series of quality cuts based on the results of the SED fitting and on the reliability of the morphological measurements.

First, we exclude sources with poor SED fits or unconstrained physical parameters, following the same approach adopted in our previous studies \citep[e.g.][]{Mountrichas2021c, Buat2021, Koutoulidis2022, Pouliasis2020}. Specifically, we require a reduced chi-square value of $\chi^2_{\rm r} \leq 5$, a threshold commonly adopted in SED-based analyses \citep[e.g.][]{Masoura2018, Buat2021}.

In addition, we require consistency between the Bayesian likelihood-weighted estimates and the best-fitting values of the M$_\star$ and SFR provided by \texttt{CIGALE}. For each parameter, \texttt{CIGALE} reports both a best-fit value, corresponding to the model with minimum $\chi^2$, and a Bayesian estimate, defined as the likelihood-weighted mean over the explored parameter space. Large discrepancies between these two estimates indicate broad or multi-peaked likelihood distributions and therefore poorly constrained physical parameters. We therefore retain only sources satisfying
\begin{equation}
\frac{1}{5} \leq \frac{\mathrm{SFR}_{\rm best}}{\mathrm{SFR}_{\rm bayes}} \leq 5
\quad \mathrm{and} \quad
\frac{1}{5} \leq \frac{M_{\star,{\rm best}}}{M_{\star,{\rm bayes}}} \leq 5 ,
\end{equation}
where the subscripts ``best'' and ``bayes'' denote the best-fit and Bayesian estimates, respectively. These criteria are identical to those adopted in our recent analyses \citep[e.g.][]{Mountrichas2021b, Mountrichas2022b, Mountrichas2022c, Buat2021, Koutoulidis2022, Pouliasis2022, Garofalo2022} and ensure that only sources with well-constrained SED-derived properties are retained. After applying the SED-fitting quality cuts, the sample is reduced to $92\,560$ galaxies.

We next assess the reliability of the morphological measurements. We consider a morphology to be reliable only if the Sérsic index, half-light radius, and ellipticity are all well defined and physically meaningful. In practice, we require finite Sérsic indices within the range $0.2 \leq n \leq 8.0$, positive and finite half-light radii, and ellipticities satisfying $e < 1$. Sources failing any of these criteria are flagged as having unreliable morphology and are excluded from analyses that explicitly rely on morphological classification.

Following these requirements, $90\,705$ galaxies, corresponding to approximately $87\%$ of the SED-cleaned sample, have reliable morphological classifications. This sample forms the basis for all subsequent analyses presented in this paper.

\subsection{Herschel far-infrared photometry}
\label{sec:herschel}

To assess the impact of far-infrared (FIR) photometric constraints on our results, we cross-match our galaxy sample with Herschel catalogues available through the Herschel Extragalactic Legacy Project (HELP), accessed via the HeDaM interface.\footnote{\url{https://hedam.lam.fr}} Specifically, we make use of the public Herschel catalogues in the GAMA09, GAMA15, and North Galactic Pole (NGP) fields, which collectively provide homogeneous FIR coverage over the DESI--eRASS1 footprint.

Approximately $45\%$ of the galaxies in our SED-cleaned sample have Herschel detections. Earlier work has demonstrated that the absence of far-infrared photometry does not significantly bias the SFR estimates obtained with \texttt{CIGALE}, provided that broad multi-wavelength coverage from the ultraviolet to the mid-infrared is available. In particular, \citet{Mountrichas2021a, Mountrichas2022a, Mountrichas2022b} showed that SFR estimates derived with and without Herschel constraints are fully consistent within the uncertainties, with no systematic offsets. 

The motivation for explicitly considering Herschel-detected and non-detected systems separately stems from our recent analysis in \citet{Mountrichas2025a}, where we found that the relation between star formation and AGN activity depends on FIR detectability. In that work (see their Fig.~8), X-ray AGN were found to exhibit systematically higher SFR than non-AGN galaxies in the non-Herschel detected population across all $M_\star$ and $L_X$ bins. In contrast, among Herschel-detected systems, this enhancement was present only at lower stellar masses ($\log[M_\star/M_\odot] < 11.0$), while at higher masses the SFRs of AGN and non-AGN galaxies became indistinguishable. 

Their analysis suggested that AGN activity is associated with enhanced star formation primarily in lower-mass galaxies, whereas the most massive systems ($\log[M_\star/M_\odot] \gtrsim 11.5$) show little difference in star-formation rates between AGN and non-AGN hosts. 
At low redshift ($z<1$), however, massive non-AGN galaxies were found to quench more rapidly than AGN hosts, while at higher redshifts ($z>1$) both populations maintained high star-formation rates, likely reflecting the larger cosmic gas supply available at earlier epochs.

Motivated by these findings, we initially split both the X-ray AGN and the non-AGN control samples in the present work into Herschel-detected and non-detected subsets, in order to test whether similar trends are recovered with the DESI--eRASS1 dataset and our updated analysis pipeline. However, as shown in Appendix~\ref{sec:app_herschel}, we find no statistically significant differences between the two subsets in the relations examined here. We therefore do not split the sample by Herschel detection in the remainder
of the analysis. In what follows, the main results are presented for the
full sample, which includes both Herschel-detected and Herschel-non-detected
sources, without applying FIR detectability as an additional selection
criterion.

We note that the absence of statistically significant differences between Herschel-detected and non-detected systems in the present analysis does not contradict our earlier findings. In \citet{Mountrichas2025a}, the separation between the two populations was partly driven by massive non-AGN galaxies with suppressed star formation, particularly at low redshift. In the present work, such systems are explicitly excluded through an sSFR-based quiescent-galaxy selection (see Sect. \ref{sec:quiescent}), which naturally reduces the contrast between FIR-detected and non-detected populations.

In addition, the eRASS1 X-ray selection probes, on average, higher AGN luminosities compared to the XMM-Newton surveys used in our previous studies \citep[e.g.][]{Brandt2015, Aird2018, Merloni2024}. This biases the sample towards more actively accreting systems, which are preferentially hosted by star-forming galaxies, largely independent of their far-infrared detectability \citep[e.g.][]{Rosario2013, Mullaney2015, Aird2019}. As a consequence, FIR-based subsamples exhibit similar SFR$_{\rm norm}$ behaviour within the parameter space explored here.

\subsection{Assessment of near-infrared photometric effects}
\label{sec:nir}

Although the DESI VAC does not uniformly include near-infrared (NIR) photometry, we assess the robustness of our SED-derived physical parameters to the absence of NIR data and quantify any resulting systematic effects. For that purpose, we cross-match our sample with the VIKING and VHS surveys \citep{Edge2013VIKING, McMahon2013VHS}, identifying $16\,558$ sources with available NIR measurements.

For these sources, we repeat the SED fitting both with and without NIR photometry. The comparison, presented in Appendix~\ref{app:nir_impact}, shows that the absence of NIR data has negligible impact on the derived SFR, while $M_\star$ and $frac_{AGN}$ exhibit systematic offsets that depend on parameter values. Based on this analysis, we apply empirical corrections to the $M_\star$ and $frac_{AGN}$ used in this work. We note that, for the results presented in Sect.~\ref{sec:results}, NIR photometry is not included in the SED fitting, even for sources with available NIR data.

\subsection{Stellar mass completeness and weighting scheme}
\label{sec:masscomp}

Any analysis linking AGN activity to host-galaxy properties requires careful treatment of $M_\star$ incompleteness, particularly in flux-limited surveys. To quantify the $M_\star$ completeness of our DESI--eRASS1 sample, we adopt the empirical approach introduced by \citet{Pozzetti2010}, which estimates the limiting M$_\star$ as a function of redshift given the survey flux limit.

The method is based on the relation
\begin{equation}
\log M_{\star,\mathrm{lim}} = \log M_\star + 0.4\,(m - m_{\mathrm{lim}}),
\end{equation}
where $m$ is the apparent magnitude of an individual galaxy, and $m_{\mathrm{lim}}$ is the limiting magnitude of the survey in the chosen band. For each redshift bin, the distribution of $M_{\star,\mathrm{lim}}$ is computed, and the M$_\star$ completeness limit is defined as the mass above which a fixed fraction (typically 90\%) of galaxies could be observed at the survey flux limit.

In this work we apply the Pozzetti et al.\ method using the DESI $z$ band. 
This band provides a good compromise between depth and sensitivity to the underlying M$_\star$ of galaxies across the redshift range considered, while remaining relatively less affected by recent star formation than bluer bands. 
We adopt an effective limiting magnitude of $z_{\mathrm{lim}} = 21.1$ AB, consistent with the typical depth of the DESI Legacy Imaging Surveys used to construct the \citet{Siudek2024} value-added catalogue. 
This value should be regarded as an effective depth used for the empirical completeness estimate rather than a strict survey selection limit.

For consistency with our previous studies of AGN host galaxies \citep[e.g.][]{Mountrichas2021c, Mountrichas2022a, Mountrichas2022b, Mountrichas2024a}, we compute completeness limits in two redshift bins, $0 \leq z < 0.5$ and $0.5 \leq z \leq 1.5$. Tests using finer redshift grids yield nearly identical limits and do not affect the results of the analysis (see also Sect.~3.3 of \citealt{Mountrichas2021c}; Sect.~3.4 of \citealt{Mountrichas2022a, Mountrichas2022b}; and Sect.~3.2.3 of \citealt{Mountrichas2024a}).

Applying this procedure, we derive M$_\star$ completeness limits of $\log(M_\star/M_\odot)=9.35$ and $\log(M_\star/M_\odot)=10.28$ for the low- and high-redshift bins, respectively. Rather than imposing strict M$_\star$ cuts at these limits, we adopt a weighting scheme to statistically correct for incompleteness. For each galaxy we compute a mass-completeness weight defined as the inverse of the fraction of galaxies observable at its M$_\star$ and redshift. These weights are propagated through all subsequent analyses involving binned statistics and median trends.

The completeness correction derived using the Pozzetti et al.\ method formally depends only on M$_\star$ and redshift. In principle, incompleteness may also depend on additional galaxy properties such as colour, star-formation history, or morphology, since galaxies with different SEDs can exhibit different mass-to-light ratios at fixed M$_\star$. However, the analyses presented in this work are restricted to the star-forming galaxy population (see Sect.~\ref{sec:quiescent}), which reduces the range of stellar populations and therefore mitigates such effects.

Importantly, the main analyses presented in this work focus on galaxies in the M$_\star$ range $10.5 \leq \log(M_\star/M_\odot) < 11.5$, which lies well above the derived mass-completeness limits in both redshift bins. Consequently, the principal results of this paper are insensitive to the precise value of the completeness limits or to the choice of photometric band used to estimate them.

As an additional robustness test, we repeat the main analyses restricting the sample to galaxies above the derived mass-completeness limits, i.e. without applying completeness weights. The resulting trends remain qualitatively unchanged, although with larger statistical uncertainties due to the reduced sample size. This confirms that the conclusions of this work are not driven by the adopted incompleteness correction scheme.

\subsection{Exclusion of quiescent systems}
\label{sec:quiescent}

\begin{figure*}
\centering
\includegraphics[height= 6.cm, width=0.8\textwidth]{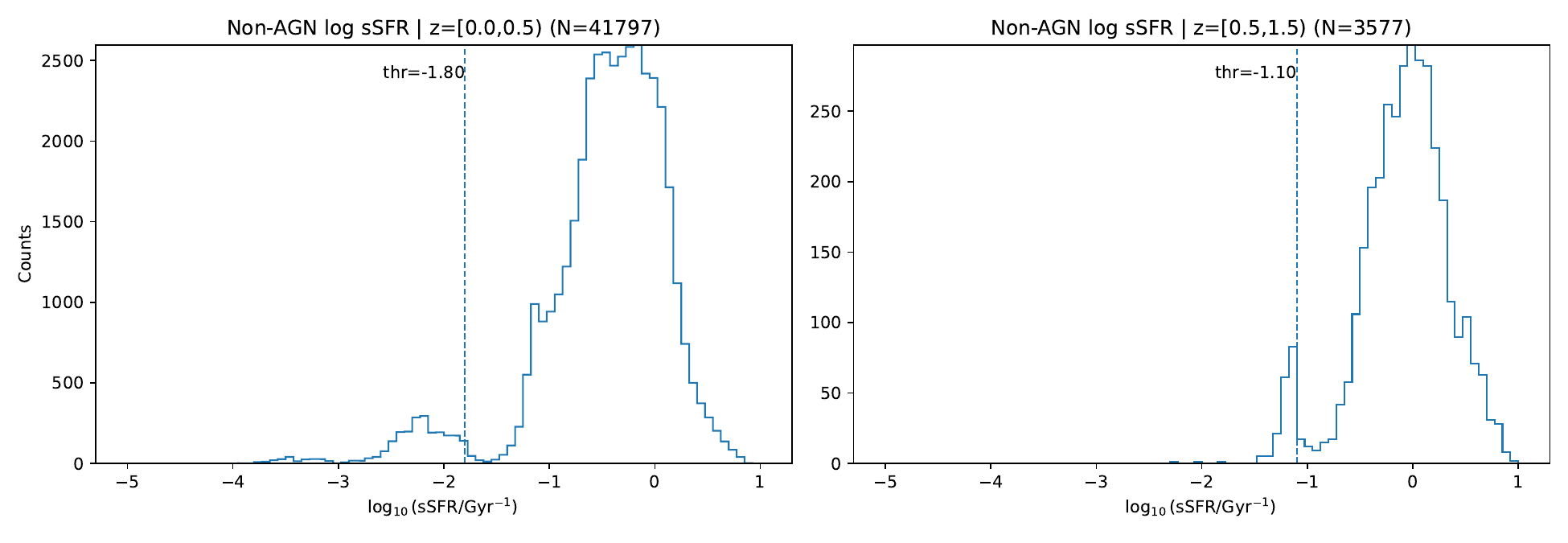}
\caption{%
Distributions of log sSFR in two redshift bins for the non-AGN population, used to define quiescent thresholds. Vertical dashed lines indicate the adopted cuts.%
}
\label{fig:ssfr_cuts}
\end{figure*}

Quiescent galaxies are excluded following the approach adopted in our previous works (e.g. Mountrichas et al. 2021c, 2022c,a, 2024a). We examine the distribution of $\log({\rm sSFR})$ for non-AGN galaxies in two redshift bins and identify the transition between the star-forming and quiescent populations using the bimodal structure of the distribution.

In practice, the $\log({\rm sSFR})$ distribution is first smoothed to reduce sensitivity to binning noise. We then identify the two dominant peaks,
corresponding to the low-sSFR quiescent population and the high-sSFR
star-forming population, while ignoring small peaks in sparsely populated
tails. The quiescent threshold is defined from the minimum between these
two peaks. The adopted thresholds are therefore intended to trace the
transition between the two physical populations, rather than the absolute minimum in the binned histogram, which can be affected by noise or by
low-number tails.

This procedure yields thresholds of
$\log({\rm sSFR}/{\rm Gyr}^{-1})=-1.8$ for $0 \leq z < 0.5$ and
$\log({\rm sSFR}/{\rm Gyr}^{-1})=-1.1$ for $0.5 \leq z \leq 1.5$.
Galaxies below these thresholds are classified as quiescent and excluded
from both the AGN and non-AGN samples used in the subsequent analysis. Approximately $16\%$ of the sources are excluded by this criterion. The sSFR distributions and adopted thresholds are shown in Fig.~\ref{fig:ssfr_cuts}. We note that small shifts of
the thresholds within the range suggested by the local distribution affect
only a small number of galaxies and do not change the results presented
below.

\subsection{Star-formation normalisation}
\label{sec:sfrnorm}

To compare the star-formation properties of AGN host galaxies with those of the general star-forming population, we compute the SFR$_{\mathrm{norm}}$ parameter \citep[e.g.,][]{Mullaney2015, Masoura2018, Masoura2021}. This quantity is defined as the ratio between the SFR of an AGN host galaxy and the SFR of non-AGN galaxies from the control sample matched in $M_\star$ and redshift. By construction, SFR$_{\mathrm{norm}}$ measures the relative enhancement or suppression of star formation in AGN hosts with respect to typical star-forming galaxies at fixed host properties.

The control sample is composed of star-forming galaxies free of both X-ray and SED-selected AGN, after applying the same redshift, SED-quality, and quiescent-galaxy selection criteria described above. Matching is performed within $\pm0.1$~dex in $M_\star$ and $\pm0.075(1+z)$ in redshift \citep[e.g.][]{Mountrichas2021c, Mountrichas2022a, Mountrichas2022b, Mountrichas2024a}. For each AGN, we compute the ratio between its SFR and the SFR of each matched control galaxy and adopt the median of this distribution as the SFR$_{\mathrm{norm}}$ value. $M_\star$ completeness weights (Sect.~\ref{sec:masscomp}) are incorporated throughout this procedure to account for residual incompleteness effects.

The control matching is performed independently for each X-ray AGN before any binning in $L_{\rm X}$, $\lambda_{\rm sBHAR}$, or morphology is applied. Thus, the SFR$_{\mathrm{norm}}$ value assigned to each AGN is always computed relative to non-AGN galaxies matched in $M_\star$ and redshift, and the plotted SFR$_{\mathrm{norm}}$ trends show the median of these individual AGN values in the corresponding bins of the analysed parameter.

To ensure statistically robust estimates of SFR$_{\mathrm{norm}}$, we exclude AGN for which fewer than 50 matched control galaxies are available. This threshold represents a compromise between minimising statistical noise in the median SFR estimate and retaining an adequate sample size. We verified that relaxing this requirement leads to larger fluctuations and increased uncertainties in the derived trends, while imposing more stringent thresholds significantly reduces the number of AGN without altering the qualitative behaviour of the results. We therefore adopt this conservative criterion for the remainder of the analysis. We also note that only bins containing more than five sources are included in the presentation of our results.

After applying all the aforementioned selection criteria, the final sample comprises $1\,171$ X-ray AGN. Of these, 417 have reliable morphological classifications (see section \ref{sec:sedmorphcuts}), including 183 hosted by disk-dominated galaxies and 188 by spheroid-dominated systems. The non-AGN control sample consists of $45\,374$ galaxies, of which $43\,066$ have reliable morphological information. Among these, $33\,177$ are classified as disk-dominated and $5\,962$ as spheroid-dominated galaxies. The approximately equal numbers of disk- and spheroid-dominated AGN hosts reflect the well-known tendency of X-ray AGN to reside preferentially in relatively massive galaxies, where both structural types are common. In contrast, the non-AGN control sample spans the full M$_\star$ range of the parent galaxy population and is therefore dominated by disk galaxies, which are more numerous at lower M$_\star$.

\section{Results}
\label{sec:results}

In this section we present the main results of this work.
We first examine the dependence of $\mathrm{SFR}_{\rm norm}$ on AGN activity, using both $L_X$ and $\lambda_{\rm sBHAR}$ as tracers. 
The quantity $\lambda_{\rm sBHAR}$ is defined as the AGN bolometric luminosity normalised by the M$_\star$ of the host galaxy, and therefore provides a proxy for the accretion rate relative to the host-galaxy mass.

While $L_X$ traces the absolute instantaneous power of the AGN, $\lambda_{\rm sBHAR}$ characterises the strength of black-hole accretion relative to the M$_\star$ of the host galaxy. 
Comparing these two quantities allows us to assess whether the connection between star formation and AGN activity is primarily linked to the overall AGN power output or to the accretion rate normalised by host-galaxy properties.

We then investigate the incidence of X-ray AGN as a function of position relative to the star-forming main sequence, and assess how these trends depend on host-galaxy morphology.

\subsection{SFR$_{\mathrm{norm}}$ as a function of $L_X$}
\label{sec:sfrnorm_lx}

\begin{figure*}[t]
\centering
\includegraphics[width=0.85\textwidth, height=10. cm]{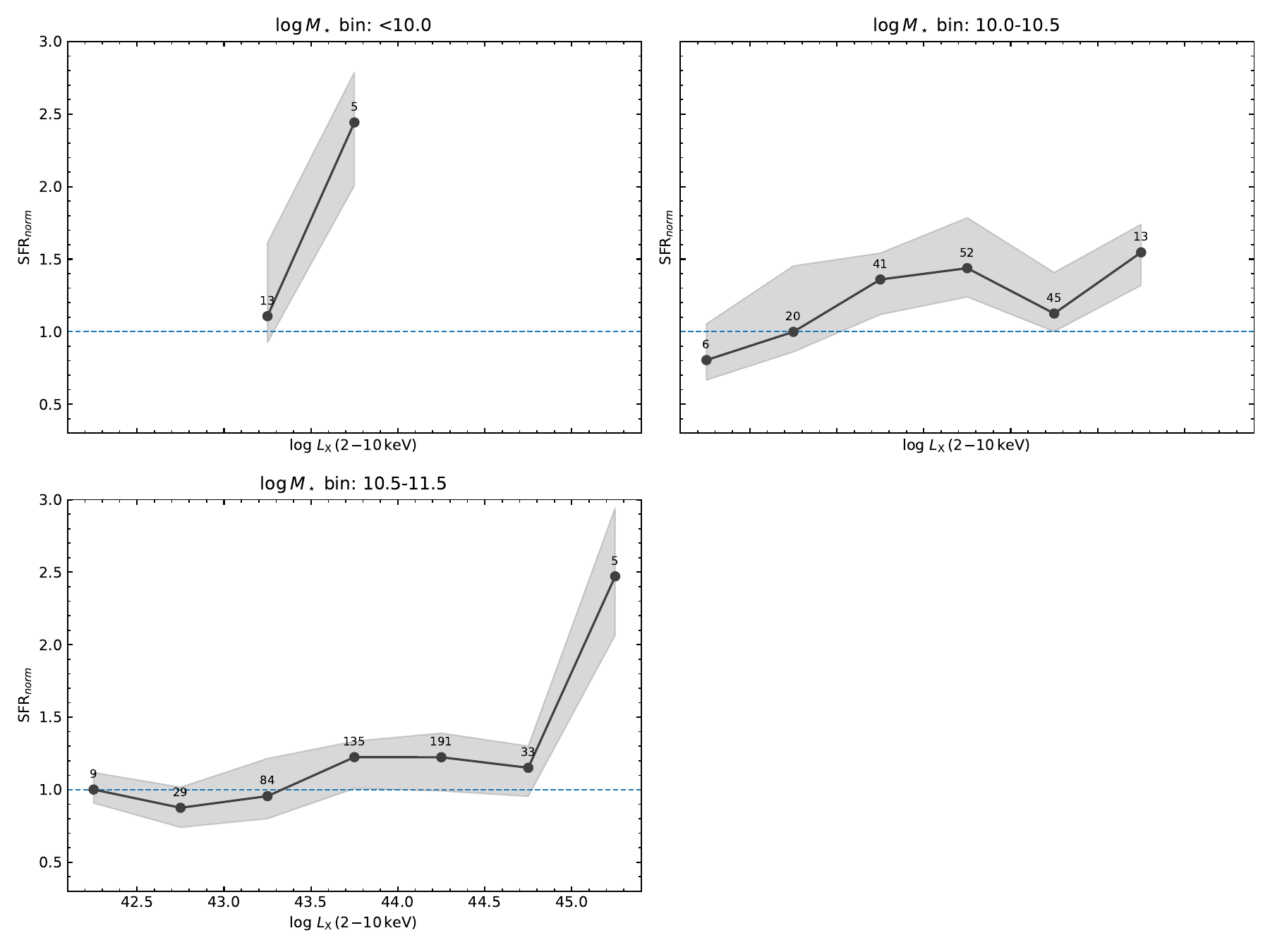}
\caption{
Normalized star formation rate ($\mathrm{SFR}_{\rm norm}$) as a function of X-ray luminosity,
$\log L_{\mathrm{X}}(2$--$10\,\mathrm{keV})$, shown in bins of stellar mass.
Each panel corresponds to a different $\log M_\star$ interval, as indicated.
Data points represent the median $\mathrm{SFR}_{\rm norm}$ in each luminosity bin, while the shaded regions show the 16th--84th percentile range for visual clarity.
The horizontal dashed line marks $\mathrm{SFR}_{\rm norm}=1$.
Numbers next to each data point indicate the number of sources contributing to that bin. For each X-ray AGN, SFRnorm is computed using non-AGN controls matched in
$M_\star$ and redshift as described in Sect.~\ref{sec:sfrnorm}; the points
show the median of the individual AGN SFRnorm values in each bin.
}
\label{fig:sfrnorm_lx_mstar_noherschel}
\end{figure*}

We begin by examining the relation between SFR$_{\mathrm{norm}}$, and $L_{\mathrm{X}}$, for the full X-ray AGN sample. This serves two purposes: first, to verify that our dataset reproduces the trends reported in previous studies; and second, to provide a baseline against which the morphology-dependent results can be compared.

As discussed in Sect.~\ref{sec:herschel}, we initially explored the SFR$_{\mathrm{norm}}$–$L_{\mathrm{X}}$ relation separately for Herschel-detected and non-detected sources, motivated by the different behaviours reported in \citet{Mountrichas2025a}. However, in the present dataset no statistically significant differences are found between the two subsets. We therefore do not distinguish between Herschel-detected and non-detected sources in the remainder of this section. The Herschel-split results are presented for completeness in Appendix~\ref{sec:app_herschel}.

Figure~\ref{fig:sfrnorm_lx_mstar_noherschel} shows $\mathrm{SFR}_{\rm norm}$ as a function of $L_{\mathrm{X}}$ in bins of $M_\star$.
The overall behaviour is broadly consistent with that reported in our previous studies (e.g. \citealt{Mountrichas2021c, Mountrichas2022a, Mountrichas2022b, Mountrichas2024a}), with $\mathrm{SFR}_{\rm norm}$ remaining close to unity at low to moderate $L_{\mathrm{X}}$ and increasing at higher $L_{\mathrm{X}}$.
As already noted in these works, the characteristic $L_{\mathrm{X}}$ at which this transition occurs shifts to higher values with increasing $M_\star$.

The limited number of AGN at the lowest $M_\star$ prevents a detailed exploration of the dwarf regime, and the trends observed in the lowest-mass bin are consistent, within uncertainties, with those reported in \citet{Mountrichas2024a} and \citet{Cristello2024}. As such, the present analysis does not significantly extend previous constraints in this regime.

\subsubsection{Dependence of the SFR$_{\mathrm{norm}}-L_X$ relation on host-galaxy morphology}
\label{sec:sfrnorm_lx_morph}

\begin{figure}[t]
\centering
\includegraphics[width=\columnwidth]{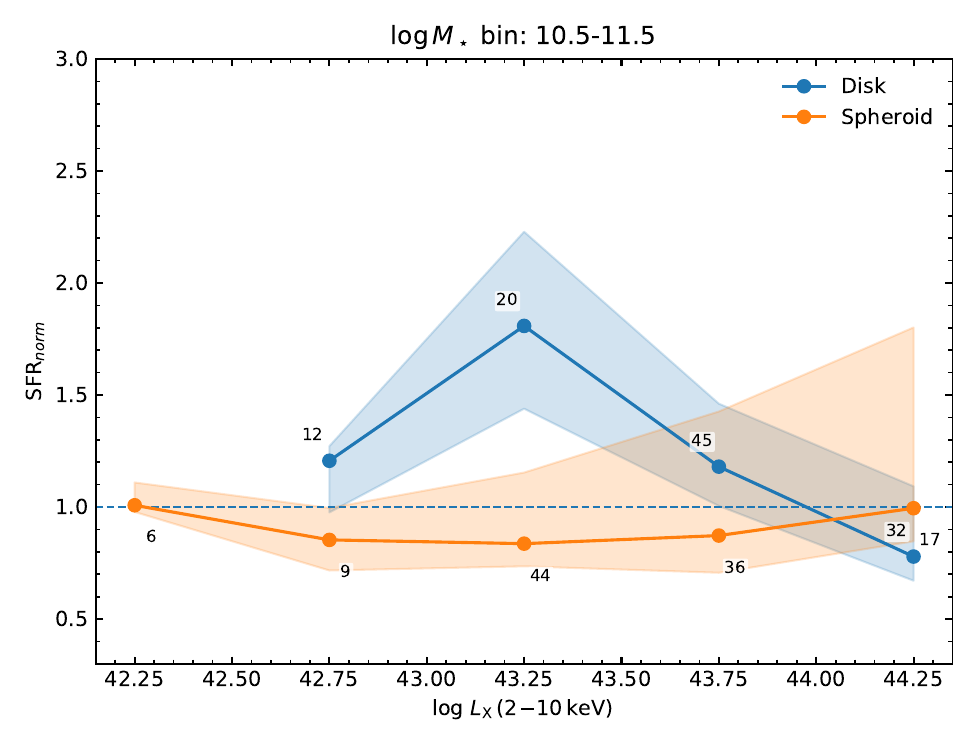}
\caption{
Normalized star formation rate ($\mathrm{SFR}_{\rm norm}$) as a function of X-ray luminosity,
$\log L_{\mathrm{X}}(2$--$10\,\mathrm{keV})$, for disk- and spheroid-dominated galaxies in the stellar mass bin
$10.5 \le \log(M_\star/{\rm M_\odot}) < 11.5$.
Points show the median $\mathrm{SFR}_{\rm norm}$ in each luminosity bin for the two morphological classes, while the shaded regions indicate the 16th--84th percentile range for visual clarity.
The horizontal dashed line marks $\mathrm{SFR}_{\rm norm}=1$.
Numbers next to the symbols indicate the number of sources contributing to each bin. For each X-ray AGN, SFRnorm is computed using non-AGN controls matched in
$M_\star$ and redshift as described in Sect.~\ref{sec:sfrnorm}; the points
show the median of the individual AGN SFRnorm values in each bin.
}
\label{fig:sfrnorm_lx_morph_overlay}
\end{figure}

We now turn to one of the main results of this work: the dependence of the SFR$_{\mathrm{norm}}$–$L_{\mathrm{X}}$ relation on host-galaxy morphology. In Fig.~\ref{fig:sfrnorm_lx_morph_overlay}, we show SFR$_{\mathrm{norm}}$ as a function of $L_{\mathrm{X}}$ separately for disk-dominated and spheroid-dominated galaxies, focusing on the $M_\star$ range $10.5 \leq \log(M_\star/M_\odot) < 11.5$, where both sufficient statistics and reliable morphological classifications are available.

A difference emerges between the two morphological classes. At low to intermediate $L_X$, disk-dominated AGN hosts exhibit systematically elevated SFR$_{\mathrm{norm}}$ values, indicating enhanced star formation relative to the matched non-AGN control sample. In contrast, spheroid-dominated AGN hosts show SFR$_{\mathrm{norm}}$ values consistent with unity across the same luminosity range, suggesting no significant enhancement relative to their non-AGN counterparts.

To quantify these trends, we fit a simple interaction model of the form
\begin{equation}
\log \mathrm{SFR}_{\mathrm{norm}} = a + b(\log L_{\mathrm{X}} - 43.5) + c\,I_{\mathrm{disk}} + d(\log L_{\mathrm{X}} - 43.5)\,I_{\mathrm{disk}},
\end{equation}
where $I_{\mathrm{disk}}$ is an indicator variable equal to unity for disk-dominated systems and zero otherwise. The best-fitting parameters are:
\begin{align*}
a &= -0.002 \pm 0.034, \\
b &= -0.003 \pm 0.056, \\
c &= +0.091 \pm 0.049, \\
d &= -0.165 \pm 0.094.
\end{align*}

The positive disk offset ($c$) is detected at the $\sim1.9\sigma$ level, while the difference in slope ($d$) is detected at the $\sim1.8\sigma$ level. Although these trends are of modest statistical significance, they consistently point towards enhanced star formation in disk-dominated AGN hosts at low and intermediate $L_{\mathrm{X}}$ compared to matched non-AGN galaxies, a behaviour not observed in spheroid-dominated systems.

These results extend the findings of \citet{Mountrichas2022c}, who reported different star-formation properties for disk- and bulge-dominated AGN hosts, by explicitly demonstrating that the divergence is most apparent when considering star formation relative to matched non-AGN galaxies. We discuss the physical implications of this behaviour in Sect.~\ref{sec:discussion}.
\subsection{SFR$_{\mathrm{norm}}$ as a function of $\lambda_{\mathrm{sBHAR}}$}
\label{sec:sfrnorm_lambda}

In addition to $L_X$, the strength of the connection between star formation and AGN activity may depend on how efficiently the black hole is accreting relative to the $M_\star$ of its host galaxy. In \citet{Mountrichas2023d}, we found that the correlation between SFR$_{\mathrm{norm}}$ and $\lambda_{\mathrm{sBHAR}}$ was generally weaker than that observed with AGN luminosity, suggesting that instantaneous accretion efficiency is not the primary driver of enhanced star formation in AGN hosts. Here we revisit this question using the DESI–eRASS1 sample, and we further explore whether any dependence on specific accretion rate differs between disk- and spheroid-dominated galaxies.

We estimate the $\lambda_{\mathrm{sBHAR}}$, following, e.g.,  Eq.~(1) of  \citet{Mountrichas2023d}, defined as
\begin{equation}
\lambda_{\mathrm{sBHAR}} = \frac{L_{\mathrm{bol}}}{1.26 \times 10^{38} \times 0.002 \, M_\star},
\end{equation}
where $L_{\mathrm{bol}}$ is the AGN bolometric luminosity derived from the SED fitting and $M_\star$ correspond to the corrected values described in Sect.~\ref{sec:masscomp}. 


\subsubsection{Global trends of SFR$_{\mathrm{norm}}-\lambda_{\mathrm{sBHAR}}$}
\label{sec:sfrnorm_lambda_all}

\begin{figure*}[t]
\centering
\includegraphics[width=0.85\textwidth, height=10. cm]{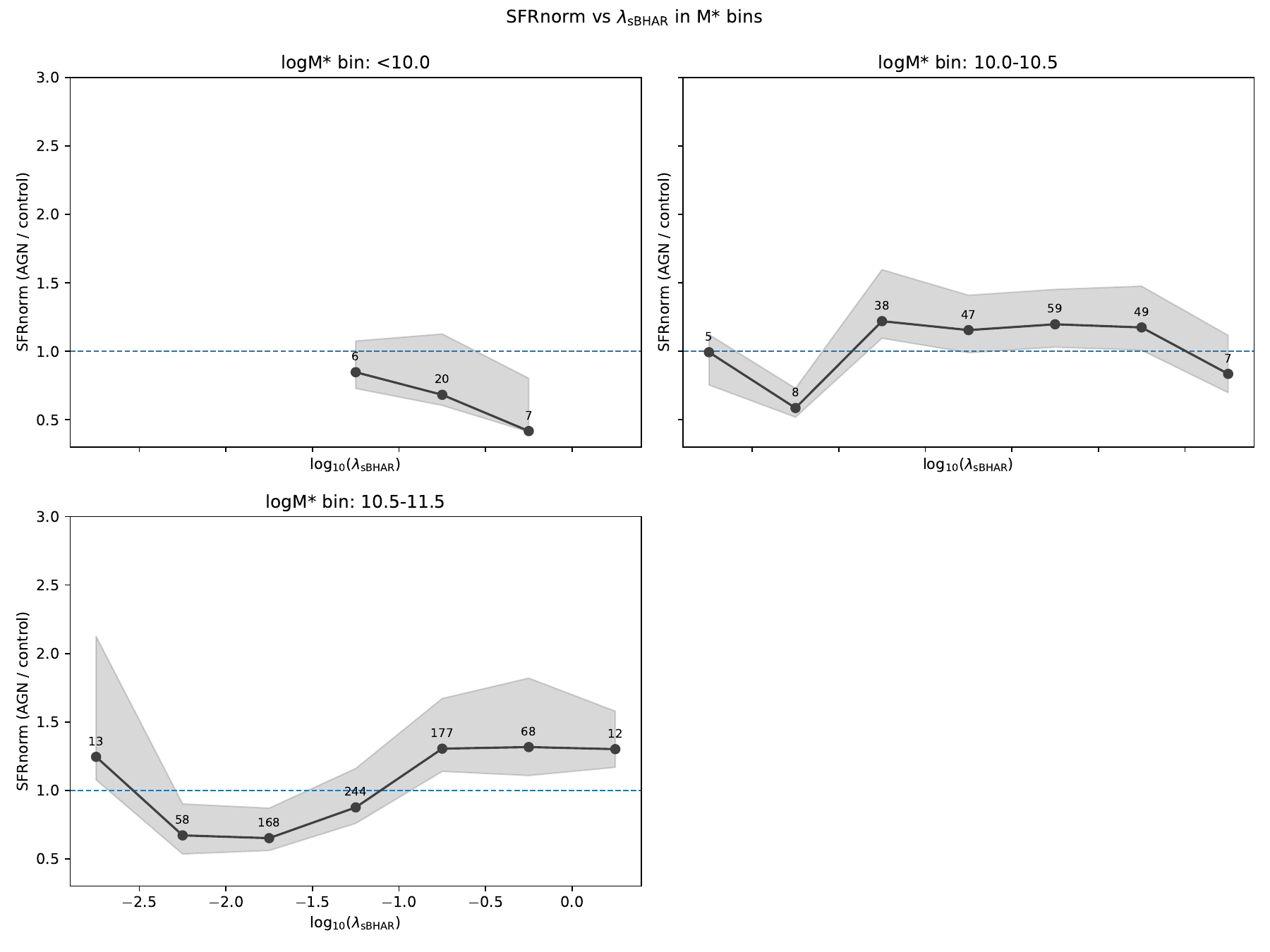}
\caption{
$\mathrm{SFR}_{\rm norm}$ as a function of $\log_{10}(\lambda_{\rm sBHAR})$, shown in bins of $M_\star$.
Each panel corresponds to a different $\log M_\star$ interval, as indicated.
Data points represent the median $\mathrm{SFR}_{\rm norm}$ in each $\lambda_{\rm sBHAR}$ bin, while the shaded regions show the 16th--84th percentile range for visual clarity.
The horizontal dashed line marks $\mathrm{SFR}_{\rm norm}=1$.
Numbers next to each data point indicate the number of sources contributing to each bin. For each X-ray AGN, SFRnorm is computed using non-AGN controls matched in
$M_\star$ and redshift as described in Sect.~\ref{sec:sfrnorm}; the points
show the median of the individual AGN SFRnorm values in each bin.
}
\label{fig:sfrnorm_lambdasbhar_mstar}
\end{figure*}

Figure~\ref{fig:sfrnorm_lambdasbhar_mstar} shows SFR$_{\mathrm{norm}}$ as a function of $\lambda_{\mathrm{sBHAR}}$ for the full X-ray AGN sample, in bins of $M_\star$. In agreement with \citet{Mountrichas2023d}, we find that SFR$_{\mathrm{norm}}$ exhibits a dependence on $\lambda_{\mathrm{sBHAR}}$ over the range probed by our data. 

Fitting a simple linear relation of the form $\log \mathrm{SFR}_{\mathrm{norm}} = a + b(\log \lambda_{\mathrm{sBHAR}} - x_0)$ yields slopes that are consistent with zero in the low- and intermediate-mass bins, while a positive slope is detected in the highest-mass bin ($10.5 \leq \log M_\star < 11.5$).
Nevertheless, the dependence of $\mathrm{SFR}_{\mathrm{norm}}$ on $\lambda_{\mathrm{sBHAR}}$ appears weaker and less systematic than the trends observed when AGN activity is traced by L$_X$.

These results reinforce the picture that the relative enhancement of star formation in AGN hosts is more closely connected to the overall AGN power output than to the instantaneous accretion efficiency normalised by stellar mass.

\subsubsection{Dependence of SFR$_{\mathrm{norm}}-\lambda_{\mathrm{sBHAR}}$ on host-galaxy morphology}
\label{sec:sfrnorm_lambda_morph}

\begin{figure}[t]
\centering
\includegraphics[width=\columnwidth]{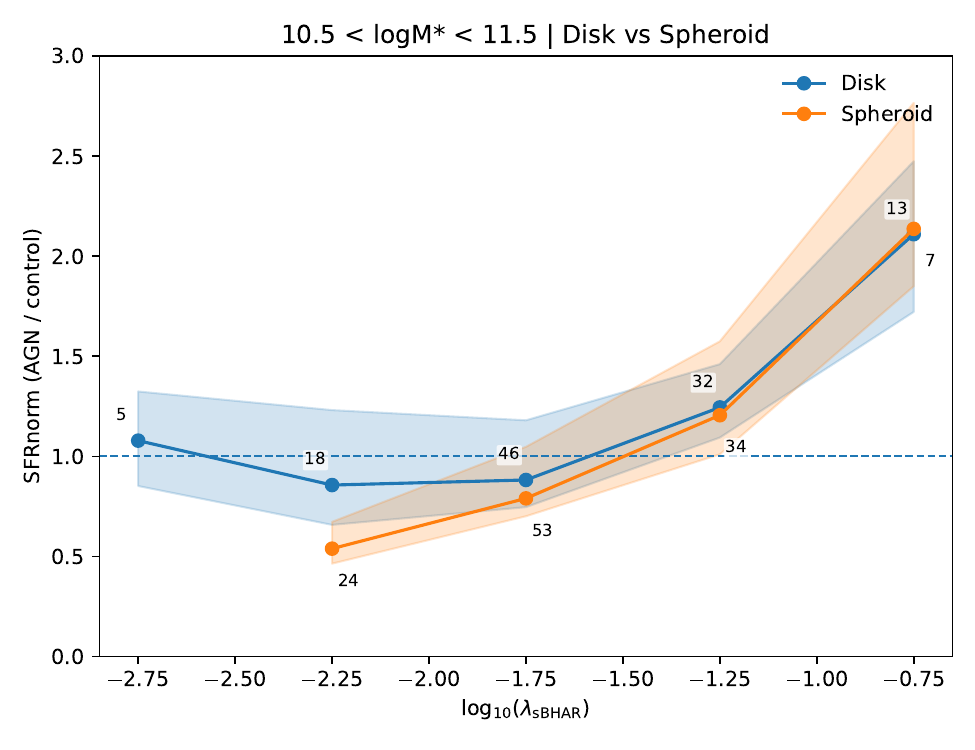}
\caption{
$\mathrm{SFR}_{\rm norm}$ as a function of $\log_{10}(\lambda_{\rm sBHAR})$, for disk- and spheroid-dominated galaxies in the stellar mass range
$10.5 \le \log(M_\star/{\rm M_\odot}) < 11.5$.
Points show the median $\mathrm{SFR}_{\rm norm}$ in each $\lambda_{\rm sBHAR}$ bin for the two morphological classes, while the shaded regions indicate the 16th--84th percentile range for visual clarity.
The horizontal dashed line marks $\mathrm{SFR}_{\rm norm}=1$.
Numbers next to the symbols indicate the number of sources contributing to each bin. For each X-ray AGN, SFRnorm is computed using non-AGN controls matched in
$M_\star$ and redshift as described in Sect.~\ref{sec:sfrnorm}; the points
show the median of the individual AGN SFRnorm values in each bin.
}
\label{fig:sfrnorm_lambdasbhar_morph_10p5_11p5}
\end{figure}

We now examine whether the relation between SFR$_{\mathrm{norm}}$ and $\lambda_{\mathrm{sBHAR}}$ depends on host-galaxy morphology. Figure~\ref{fig:sfrnorm_lambdasbhar_morph_10p5_11p5} shows SFR$_{\mathrm{norm}}$ as a function of $\lambda_{\mathrm{sBHAR}}$ separately for disk-dominated and spheroid-dominated AGN hosts,  in the $M_\star$ range $10.5 \leq \log(M_\star/M_\odot) < 11.5$, where the sample size is sufficient to robustly characterise trends.

Unlike the case of $L_X$ (Sect.~\ref{sec:sfrnorm_lx_morph}), no morphological separation is observed. Disk- and spheroid-dominated systems follow similar trends, with SFR$_{\mathrm{norm}}$ remaining close to unity across the full range of $\lambda_{\mathrm{sBHAR}}$, with the exception of the highest $\lambda_{\mathrm{sBHAR}}$ bins for both morphological types. We quantify this behaviour using the same interaction model adopted in Sect.~\ref{sec:sfrnorm_lx_morph}, replacing $\log L_{\mathrm{X}}$ with $\log \lambda_{\mathrm{sBHAR}}$. The best-fitting parameters are:
\begin{align*}
a &= +0.010 \pm 0.030, \\
b &= +0.135 \pm 0.073, \\
c &= +0.033 \pm 0.046, \\
d &= +0.028 \pm 0.104.
\end{align*}

Both the disk offset ($c$) and the slope difference ($d$) are consistent with zero, with significances below $1\sigma$. This indicates that $\lambda_{\mathrm{sBHAR}}$ does not encode a strong or morphology-dependent connection between AGN activity and host-galaxy star formation.

Taken together, these results suggest that the enhanced star formation observed in disk-dominated AGN hosts at low and intermediate $L_{\mathrm{X}}$ is not primarily driven by variations in accretion efficiency. Instead, it is likely linked to larger-scale gas availability or fuelling processes that simultaneously sustain star formation and moderate AGN activity, a scenario we discuss further in Sect.~\ref{sec:discussion}.

\subsection{Incidence of X-ray AGN across the star-forming main sequence}
\label{sec:incidence}

The analyses presented in Sects.~\ref{sec:sfrnorm_lx} and \ref{sec:sfrnorm_lambda} focus on the star-formation properties of galaxies hosting X-ray AGN, relative to carefully matched non-AGN control samples. A complementary and equally important perspective is provided by the incidence of AGN activity itself, that is, the probability that a star-forming galaxy hosts an X-ray AGN as a function of its position relative to the star-forming main sequence.

Unlike SFR$_{\mathrm{norm}}$, which probes the average enhancement or suppression of star formation in AGN hosts, AGN incidence directly quantifies how frequently nuclear activity occurs in galaxies with given star-formation properties. This makes it a particularly powerful diagnostic of the connection between star formation and black-hole growth.

\subsubsection{Definition of the star-forming main sequence offset}
\label{sec:deltams}

We characterise the position of galaxies relative to the star-forming main sequence using the quantity
\begin{equation}
\Delta \mathrm{MS} = \log \mathrm{SFR} - \log \mathrm{SFR}_{\mathrm{MS}}(M_\star, z),
\end{equation}
where $\mathrm{SFR}_{\mathrm{MS}}(M_\star, z)$ is the median star-formation rate of non-AGN, star-forming galaxies at a given $M_\star$ and redshift. The main sequence is constructed separately in two redshift intervals ($0 \leq z < 0.5$ and $0.5 \leq z \leq 1.5$), using the DESI galaxy sample after excluding quiescent systems as described in Sect.~\ref{sec:quiescent}. 

Only galaxies within the $M_\star$ range $10.5 \leq \log(M_\star/M_\odot) < 11.5$ are considered, ensuring high mass completeness in both redshift bins. The resulting $\Delta \mathrm{MS}$ values are used consistently for both AGN and non-AGN populations.

Although quiescent galaxies are excluded from the analysis, star-forming galaxies still span a substantial range of offsets around the main sequence. The parameter $\Delta{\rm MS}$ therefore measures whether a galaxy lies below, on, or above the typical star-forming sequence at its M$_\star$ and redshift. This quantity provides a convenient way to distinguish between moderately star-forming galaxies and systems undergoing enhanced star-formation episodes (often referred to as starbursts). Studying AGN incidence as a function of $\Delta{\rm MS}$ therefore allows us to test whether black-hole growth is preferentially associated with elevated star-formation states within the star-forming population.

\subsubsection{AGN incidence as a function of $\Delta$MS}
\label{sec:incidence_all}

\begin{figure*}[t]
\centering
\includegraphics[width=\textwidth]{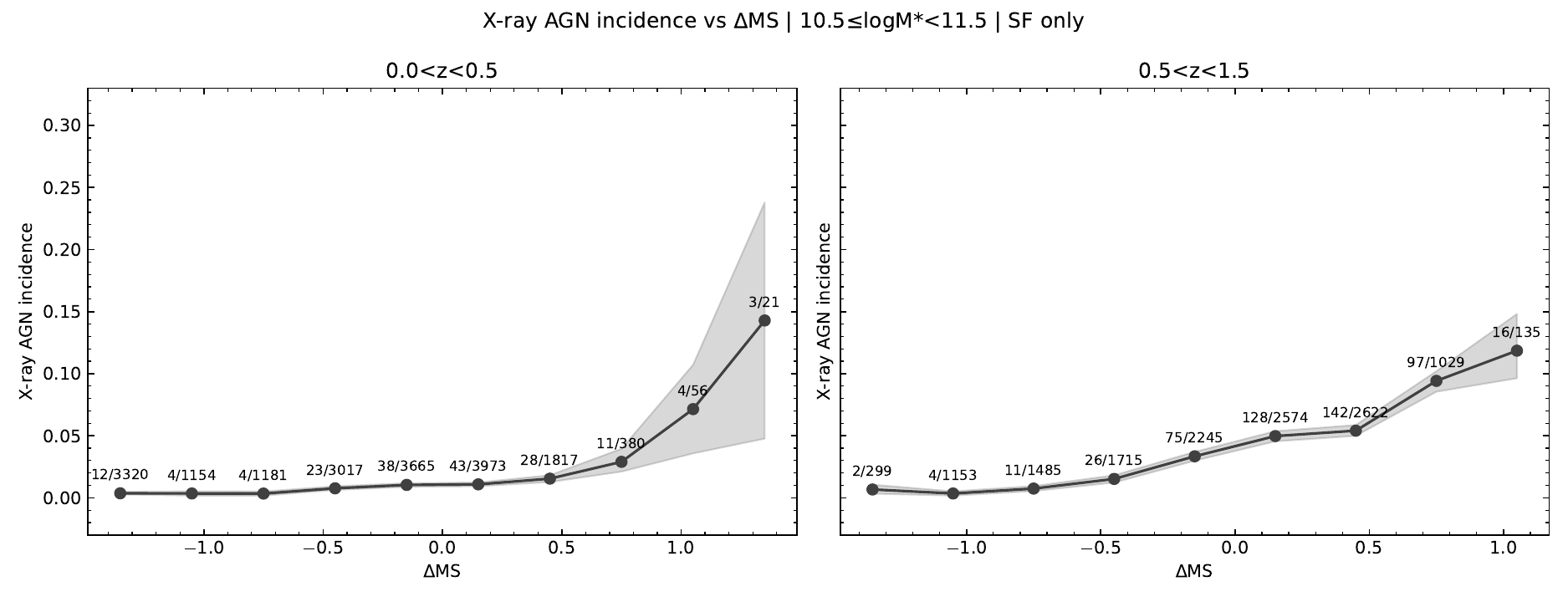}
\caption{X-ray AGN incidence as a function of the offset from the star-forming main sequence, $\Delta{\rm MS}$, for star-forming galaxies, in the stellar mass range
$10.5 \le \log(M_\star/{\rm M_\odot}) < 11.5$.
The left and right panels correspond to $0.0<z<0.5$ and $0.5<z<1.5$, respectively.
Shaded regions represent binomial uncertainties on the AGN incidence.
Numbers next to each data point indicate the number of AGN hosts and the total number of galaxies in each bin.
}
\label{fig:agn_incidence_deltams_all}
\end{figure*}

Figure~\ref{fig:agn_incidence_deltams_all} shows the fraction of galaxies hosting an X-ray AGN as a function of $\Delta \mathrm{MS}$, computed as the ratio of X-ray AGN to the total number of galaxies in each bin. Mass-completeness weights are applied consistently to both AGN and non-AGN populations.

In both redshift bins, the AGN incidence increases strongly with $\Delta \mathrm{MS}$, indicating that galaxies with elevated SFR relative to the main sequence are significantly more likely to host X-ray AGN. This trend is well described by a model of the form
\begin{equation}
\mathrm{logit}(p) = a + b\,\Delta \mathrm{MS},
\end{equation}
where $\mathrm{logit}(p)=\ln[p/(1-p)]$ is the log-odds transformation of the probability $p$ that a galaxy hosts an X-ray AGN.

The best-fitting parameters are:
\begin{itemize}
\item $0 \leq z < 0.5$: $a = -4.454 \pm 0.076$, $b = +0.802 \pm 0.118$,
\item $0.5 \leq z \leq 1.5$: $a = -3.424 \pm 0.052$, $b = +1.645 \pm 0.096$.
\end{itemize}

The positive slopes indicate that AGN incidence rises steeply with increasing $\Delta \mathrm{MS}$ in both redshift intervals, with a stronger dependence observed at higher redshift.

\subsubsection{Dependence on host-galaxy morphology}
\label{sec:incidence_morph}

\begin{figure*}[t]
\centering
\includegraphics[width=\textwidth]{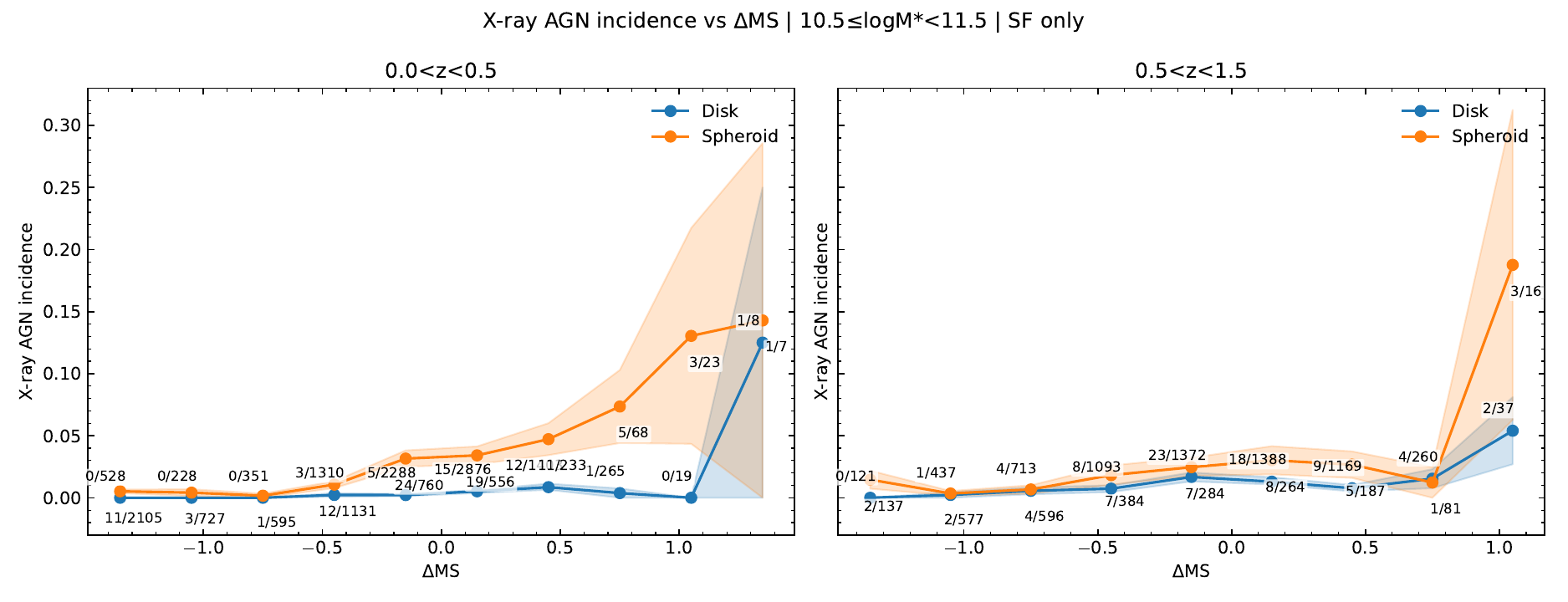}
\caption{
X-ray AGN incidence as a function of distance from the star-forming main sequence,
$\Delta{\rm MS} = \log({\rm SFR}) - \log({\rm SFR}_{\rm MS})$,
for star-forming galaxies in the stellar mass range
$10.5 \le \log(M_\star/{\rm M_\odot}) < 11.5$.
The two panels correspond to the redshift intervals $0.0<z<0.5$ (left) and $0.5<z<1.5$ (right).
Blue and orange symbols show disk- and spheroid-dominated systems, respectively, using only sources with reliable morphological classifications.
Shaded regions indicate binomial uncertainties on the AGN incidence.
Numbers next to each data point denote the number of AGN hosts and the total number of galaxies in each bin.
}
\label{fig:agn_incidence_deltams_morph}
\end{figure*}

We next investigate whether the incidence of X-ray AGN across the main sequence depends on host-galaxy morphology. Figure~\ref{fig:agn_incidence_deltams_morph} shows the AGN incidence as a function of $\Delta \mathrm{MS}$ separately for disk-dominated and spheroid-dominated galaxies.

To quantify morphological differences, we fit an interaction model of the form
\begin{equation}
\mathrm{logit}(p) = a + b\,\Delta \mathrm{MS} + c\,I_{\mathrm{disk}} + d\,\Delta \mathrm{MS}\,I_{\mathrm{disk}},
\end{equation}
where $I_{\mathrm{disk}}$ is an indicator variable equal to unity for disk-dominated galaxies and zero for spheroid-dominated systems.

In the low-redshift bin ($0 \leq z < 0.5$), the best-fitting parameters are:
\begin{align*}
a &= -3.582 \pm 0.110, \\
b &= +1.217 \pm 0.140, \\
c &= -1.995 \pm 0.210, \\
d &= +0.459 \pm 0.431.
\end{align*}

A likelihood-ratio test comparing the interaction model to the morphology-independent model yields $2\Delta\ln\mathcal{L} = 107.7$ for two additional degrees of freedom, corresponding to a probability $p = 4.1 \times 10^{-24}$. This indicates that host-galaxy morphology provides a highly significant improvement in describing the AGN incidence at fixed $\Delta \mathrm{MS}$.

In the higher-redshift bin ($0.5 \leq z \leq 1.5$), we obtain:
\begin{align*}
a &= -3.861 \pm 0.165, \\
b &= +1.476 \pm 0.216, \\
c &= -0.681 \pm 0.204, \\
d &= -0.597 \pm 0.312,
\end{align*}
with a likelihood-ratio test yielding $2\Delta\ln\mathcal{L} = 12.5$ and $p = 1.9 \times 10^{-3}$. Although weaker than at low redshift, the dependence of AGN incidence on morphology remains statistically significant.

Overall, these results demonstrate that the dependence of X-ray AGN incidence on $\Delta \mathrm{MS}$ is modulated by host-galaxy morphology, with a statistically significant difference between disk- and spheroid-dominated systems in both redshift intervals.
Interestingly, the relative behaviour of the two morphological classes differs with redshift: at low redshift ($0 \leq z < 0.5$), spheroid-dominated galaxies show a higher AGN incidence at elevated $\Delta \mathrm{MS}$, while at higher redshift ($0.5 \leq z \leq 1.5$) disk-dominated systems exhibit comparable or higher AGN incidence.
This suggests that the interplay between star formation and black-hole growth depends not only on the level of star formation relative to the main sequence, but also on both the structural properties of the host galaxy and cosmic epoch.

\section{Discussion}
\label{sec:discussion}

In this section, we place our findings in the context of previous studies and discuss their physical implications. In particular, we focus on the role of host-galaxy morphology and the incidence of AGN activity across the star-forming main sequence.

The novelty of the present work is not simply the measurement of
SFR$_{\mathrm{norm}}$ or AGN incidence separately, but their combination
within the same DESI--eRASS1 framework and their joint interpretation as a function of host-galaxy morphology. SFR$_{\mathrm{norm}}$ addresses whether AGN hosts have enhanced or suppressed star formation relative to matched
non-AGN galaxies at fixed $M_\star$ and redshift, while the incidence
analysis quantifies how the probability of hosting an X-ray AGN varies
across the star-forming main sequence. Combining these two diagnostics
allows us to distinguish between enhanced star formation in AGN hosts and
an increased probability of triggering AGN activity in galaxies with
elevated SFR.

\subsection{Star formation enhancement and the role of morphology}

The results presented in Sect.~\ref{sec:sfrnorm_lx} confirm earlier findings that the star formation of X-ray AGN relative to non-AGN galaxies matched in $M_\star$ and redshift depends on both $L_{\rm X}$ and $M_\star$ \citep[e.g.,][]{Mountrichas2021c, Mountrichas2022a, Mountrichas2022b, Mountrichas2024a, Cristello2024}. 
As shown in Fig.~\ref{fig:sfrnorm_lx_mstar_noherschel}, $\mathrm{SFR}_{\rm norm}$ remains close to unity at low to intermediate $L_{\rm X}$ and increases at higher luminosities, with the characteristic transition shifting toward higher $L_{\rm X}$ in more massive systems.

These results are broadly consistent with studies showing that
X-ray selected AGN are commonly hosted by star-forming galaxies and that
the connection between star formation and black-hole growth depends on
both AGN luminosity and host-galaxy properties
(e.g. \citealt{Rosario2013,Mullaney2015,Aird2018,Aird2019}). In this
context, the increase of SFR$_{\mathrm{norm}}$ at high $L_{\rm X}$ can be
interpreted as evidence that the most luminous X-ray AGN preferentially
occur in systems with enhanced cold-gas reservoirs. However, our results
also show that this behaviour is not uniform across galaxy structure. The
enhancement at moderate $L_{\rm X}$ is primarily associated with
disk-dominated systems, suggesting that the availability of gas on
galaxy-wide scales, rather than accretion efficiency alone, plays an
important role in linking star formation and AGN activity.

A key new result of this work is that this behaviour depends on host-galaxy morphology. In the $M_\star$ range $10.5 \le \log(M_\star/{\rm M_\odot}) < 11.5$, disk-dominated AGN hosts exhibit enhanced $\mathrm{SFR}_{\rm norm}$ at low to intermediate $L_{\rm X}$, whereas spheroid-dominated systems remain consistent with unity. 
This indicates that the enhanced star formation observed in moderate-luminosity AGN is primarily associated with disk-dominated systems.

This interpretation is also consistent with morphological studies showing
that moderate-luminosity AGN are frequently found in disk galaxies and do
not necessarily require major mergers as their dominant triggering
mechanism (e.g. \citealt{Cisternas2011,Kocevski2012,Simmons2013}). In
such systems, secular processes, disk instabilities, or minor interactions
may supply gas to the central regions while also sustaining elevated star
formation. In contrast, spheroid-dominated galaxies may have more centrally
concentrated stellar distributions and lower cold-gas fractions, so that
AGN activity can occur without a strong galaxy-wide SFR enhancement.

In \citet{Mountrichas2022c}, differences in stellar population ages between bulge- and non-bulge-dominated AGN hosts were reported based on D4000 measurements (see Fig.~16 therein). 
Bulge-dominated AGN hosts exhibited, on average, younger stellar populations than their non-AGN counterparts, while non-bulge-dominated AGN hosts showed the opposite trend. 
The present analysis, which focuses on instantaneous SFR relative to matched control samples, instead reveals enhanced SFR$_{\rm norm}$ in disk-dominated systems and $\mathrm{SFR}_{\rm norm}\approx1$ in spheroid systems (Fig.~\ref{fig:sfrnorm_lx_morph_overlay}). 
This difference likely indicates the distinct timescales probed by the two diagnostics: D4000 traces stellar populations integrated over $\sim$Gyr timescales, whereas SFR$_{\rm norm}$ is sensitive to more recent ($\lesssim100$ Myr) star-formation activity.

The morphological classifications adopted in the two studies are conceptually similar but not identical. 
In \citet{Mountrichas2022c}, we used the bulge/non-bulge classifications provided by \citet{Ni2021}, based on multi-component structural decompositions of galaxy light profiles. 
In the present work, morphology is derived from DESI structural parameters using Sérsic-index thresholds to define disk- and spheroid-dominated systems. 
Although bulge-dominated systems broadly correspond to the spheroid class adopted here, and non-bulge-dominated systems to disk-dominated galaxies, the exact definitions and selection criteria differ, which should be borne in mind when comparing the detailed trends.

The absence of a comparable morphological separation when using $\lambda_{\rm sBHAR}$ (Fig.~\ref{fig:sfrnorm_lambdasbhar_morph_10p5_11p5}) is particularly informative. 
While $L_{\rm X}$ shows a morphology-dependent link with star formation, $\lambda_{\rm sBHAR}$ does not. 
This suggests that the observed enhancement in disk-dominated hosts is not primarily driven by variations in instantaneous accretion efficiency, but rather by the overall gas supply. 
In gas-rich disk galaxies, large-scale reservoirs can simultaneously sustain elevated star formation and moderate AGN luminosities. 
In contrast, spheroid-dominated systems, which are typically more gas-poor or dynamically stabilised, do not show comparable star-formation enhancement at fixed $L_{\rm X}$.

\subsection{AGN incidence across the main sequence}

The strong increase of AGN incidence with $\Delta$MS provides an important complement to the SFR$_{\rm norm}$ analysis. 
Galaxies located above the star-forming main sequence are significantly more likely to host an X-ray AGN, indicating that elevated star-formation states are closely linked to black-hole growth.

This behaviour is broadly consistent with the framework proposed by \citet{Aird2018, Aird2019}, in which AGN accretion traces the availability of cold gas in star-forming galaxies. 
However, when combined with our previous findings \citep[e.g.,][]{Mountrichas2024a}, a more nuanced picture emerges. In that paper we found that AGN-hosting galaxies follow a shallower SFR–$M_\star$ relation than non-AGN systems, implying that the increase of SFR with $M_\star$ is less pronounced in AGN hosts.

When considered jointly, these results suggest that AGN are preferentially triggered in galaxies experiencing relative SFR enhancement (high $\Delta$MS), yet the scaling of star formation with $M_\star$ differs from that of the overall galaxy population. 
A plausible interpretation is that while elevated gas reservoirs increase both SFR and the probability of AGN triggering, the partitioning of this gas between large-scale star formation and nuclear accretion differs from that in non-AGN galaxies. 
This naturally accounts for both the rising AGN incidence with $\Delta$MS and the shallower SFR–$M_\star$ slope observed previously.

\subsection{Limitations and caveats}

Several limitations should be kept in mind when interpreting these
results. First, the morphology classification is based on single-Sérsic
indices and therefore provides a broad separation between disk- and
spheroid-dominated systems, rather than a full bulge--disk decomposition.
This approach is well suited to large statistical samples, but it cannot
capture more detailed structural features such as bars, tidal
disturbances, or compact bulges. Second, the morphology-dependent analysis
is restricted mainly to the stellar-mass range
$10.5 \leq \log(M_\star/M_\odot) < 11.5$, where both the AGN statistics
and the morphological classifications are sufficiently reliable. The
present data therefore do not allow a comprehensive investigation of the
lower-mass regime, $\log(M_\star/M_\odot) < 10.5$. Finally, the SFRs
are derived from SED fitting and trace star formation over timescales that
are generally longer than the instantaneous X-ray variability of the AGN.
These limitations do not affect the main qualitative trends, but they
should be considered when interpreting the results physically.

\subsection{A unified picture}

Taken together, the SFR$_{\rm norm}$ and AGN incidence results point toward a scenario in which the connection between star formation and AGN activity is governed primarily by global gas availability, but modulated by host-galaxy structure and cosmic epoch.

Disk-dominated systems sustain enhanced star formation at fixed $L_{\rm X}$, consistent with long-lived, gas-rich fuelling modes that couple large-scale star formation and moderate AGN activity. 
Spheroid-dominated systems show weaker or absent star-formation enhancement at fixed luminosity, but can exhibit elevated AGN incidence during episodes of large $\Delta{\rm MS}$, particularly at low redshift.

Within this framework, $L_{\rm X}$ traces the overall power output linked to gas supply, whereas $\lambda_{\rm sBHAR}$ reflects accretion efficiency normalised by M$_\star$ and does not appear to encode a strong connection to host-wide star formation. 
The differing behaviour between morphological classes and redshift intervals suggests that AGN fuelling mechanisms are not universal, but depend on both structural properties and evolutionary stage.

Future constraints on cold gas content, merger incidence, and environment will be essential to determine whether the observed trends are primarily driven by secular disk instabilities, merger-induced inflows, or morphological quenching processes.

\section{Summary}
\label{sec:summary}
In this work, we investigate the star-formation properties and incidence of X-ray AGN across the star-forming main sequence using the DESI–eRASS1 dataset. The analysis includes 1,171 X-ray selected AGN (417 with reliable morphological classifications) and 45,374 non-AGN galaxies (43,066 with reliable morphology), spanning redshifts $z \leq 1.5$.

We quantified star formation in AGN hosts relative to carefully matched control samples using the SFR$_{\rm norm}$ parameter, and examined AGN activity as a function of both $L_{\rm X}$ and $\lambda_{\rm sBHAR}$. We further investigated the incidence of X-ray AGN as a function of distance from the star-forming main sequence ($\Delta{\rm MS}$) and explored the role of host-galaxy morphology. 

Special attention was given to potential systematic effects. We explicitly tested the impact of missing NIR photometry on $M_\star$, SFR and $f_{AGN}$ estimates, and applied empirical corrections where necessary. We also examined the influence of far-infrared (Herschel) photometry and found no statistically significant differences between Herschel-detected and non-detected subsamples in the present dataset.

Our main results can be summarised as follows:

\begin{enumerate}

\item SFR$_{\rm norm}$ remains close to unity at low to intermediate $L_{\rm X}$ and increases at higher luminosities, with the transition shifting toward higher $L_{\rm X}$ in more massive systems, consistent with previous studies.

\item The SFR$_{\rm norm}$–$L_{\rm X}$ relation depends on host morphology. In the $M_\star$ range $10.5 \le \log(M_\star/{\rm M_\odot}) < 11.5$, disk-dominated AGN hosts exhibit enhanced SFR$_{\rm norm}$ at low to intermediate $L_{\rm X}$, whereas spheroid-dominated systems remain consistent with unity.

\item No comparable morphology dependence is observed when AGN activity is traced by $\lambda_{\rm sBHAR}$, indicating that the star-formation enhancement in disk-dominated hosts is more closely linked to global gas supply than to instantaneous accretion efficiency.

\item The incidence of X-ray AGN increases strongly with $\Delta{\rm MS}$ in both redshift bins, with a steeper dependence at $0.5 \le z \le 1.5$, indicating that galaxies above the main sequence are significantly more likely to host an AGN.

\item The dependence of AGN incidence on $\Delta{\rm MS}$ is modulated by morphology in a redshift-dependent manner. At low redshift, spheroid-dominated systems exhibit higher AGN incidence at large $\Delta{\rm MS}$, while at higher redshift disk-dominated galaxies show comparable or higher incidence. This behaviour suggests an evolution in the dominant fuelling mechanisms of AGN activity.

\end{enumerate}

Overall, our results indicate that the connection between star formation and AGN activity is primarily governed by the availability of cold gas, but regulated by host-galaxy structure and cosmic epoch. The differing behaviour observed between $L_{\rm X}$ and $\lambda_{\rm sBHAR}$ further suggests that absolute AGN power is more directly coupled to host-wide star formation than accretion efficiency normalised by stellar mass.

Future wide-field surveys such as \textit{Euclid} \citep{Laureijs2011} will dramatically enhance this type of analysis. With its high-resolution near-infrared imaging and homogeneous morphological measurements over large cosmological volumes, \textit{Euclid} will enable more precise structural classifications and robust constraints on the evolution of the AGN–host connection across cosmic time. Combined with ongoing and future X-ray surveys, this will provide unprecedented insight into the fuelling and co-evolution of galaxies and their central black holes.

\begin{acknowledgements}
GM acknowledges funding from grant PID2021-122955OB-C41 funded by MCIN/AEI/10.13039/501100011033 and by “ERDF/EU”. This work was partially supported by the European Union's Horizon 2020 Research and Innovation program under the Maria Sklodowska-Curie grant agreement (No. 754510). This publication is part of the R\&D\&I project PID2024-155779OB-C31, funded by MICIU/AEI/10.13039/501100011033 and co-funded by FEDER, EU.

\end{acknowledgements}

\bibliography{mybib}
\bibliographystyle{aa}

\appendix

\section{Validation of the Sérsic-based morphological classification}
\label{sec:appendix_morphology}

\begin{figure}[t]
\centering
\includegraphics[width=\columnwidth]{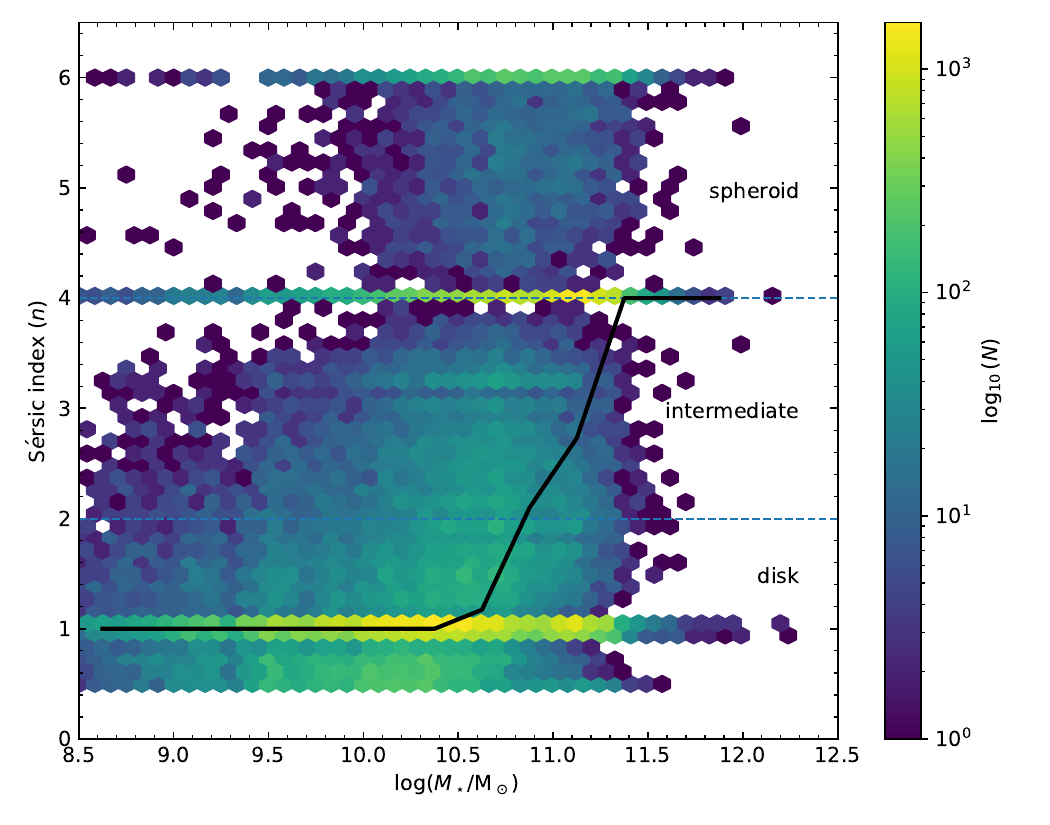}
\caption{
Distribution of Sérsic index ($n$) as a function of stellar mass for galaxies with reliable morphological measurements in the DESI--eRASS1 sample. 
The colour scale indicates the logarithm of the number of galaxies in each bin. 
The dashed horizontal lines mark the thresholds adopted in this work to define disk-dominated ($n<2$), intermediate ($2\leq n<4$), and spheroid-dominated ($n\geq4$) systems. 
The solid line shows the running median Sérsic index as a function of stellar mass, highlighting the tendency for more massive galaxies to exhibit higher central concentration.
}
\label{fig:sersic_mstar}
\end{figure}

Figure~\ref{fig:sersic_mstar} shows the distribution of Sérsic index ($n$) as a function of M$_\star$ for galaxies with reliable morphological measurements in the DESI--eRASS1 parent sample. 
The colour scale represents the logarithm of the number of galaxies per bin.

As expected, the majority of galaxies populate the low-$n$ regime ($n\sim1$), corresponding to disk-dominated systems, while a secondary population at higher Sérsic indices ($n\gtrsim4$) corresponds to spheroid-dominated galaxies. 
The dashed horizontal lines indicate the thresholds adopted in this work to define disk-dominated ($n<2$), intermediate ($2\leq n<4$), and spheroid-dominated ($n\geq4$) systems.

The solid curve shows the running median Sérsic index as a function of M$_\star$. 
This relation illustrates the well-known structural trend in which more massive galaxies tend to exhibit higher central concentration and larger bulge components (e.g. \citealt{vanDerWel2014,Bluck2014}).

Overall, the figure demonstrates that the adopted Sérsic-index thresholds isolate distinct structural regimes within the galaxy population and therefore provide a practical first-order separation between disk- and spheroid-dominated systems in large statistical samples.

\section{Impact of Herschel photometry on the SFR$_{\mathrm{norm}}$--$L_X$ relation}
\label{sec:app_herschel}

\begin{figure*}
\centering
\includegraphics[height= 6.5cm, width=\textwidth]{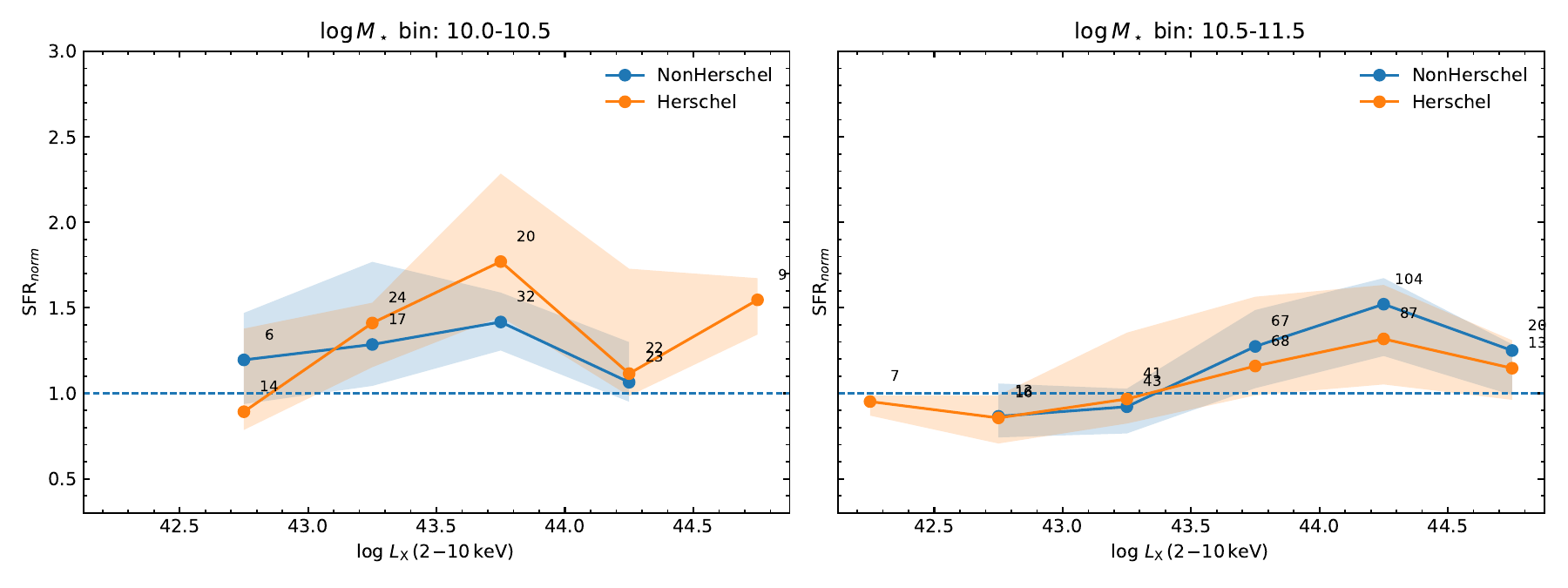}
\caption{Median SFR$_{norm}$ as a function of $L_{\mathrm{X}}$ is shown for two stellar mass bins. Blue and orange symbols correspond to non-Herschel-detected and Herschel-detected sources, respectively. Shaded areas indicate the $16$th–$84$th percentile range, and numbers denote the number of AGN per bin. The dashed line marks SFR$_{\mathrm{norm}} = 1$.}
\label{fig:sfrnorm_lx_herschel}
\end{figure*}

\begin{figure*}
\centering
\includegraphics[height= 6.5cm, width=\textwidth]{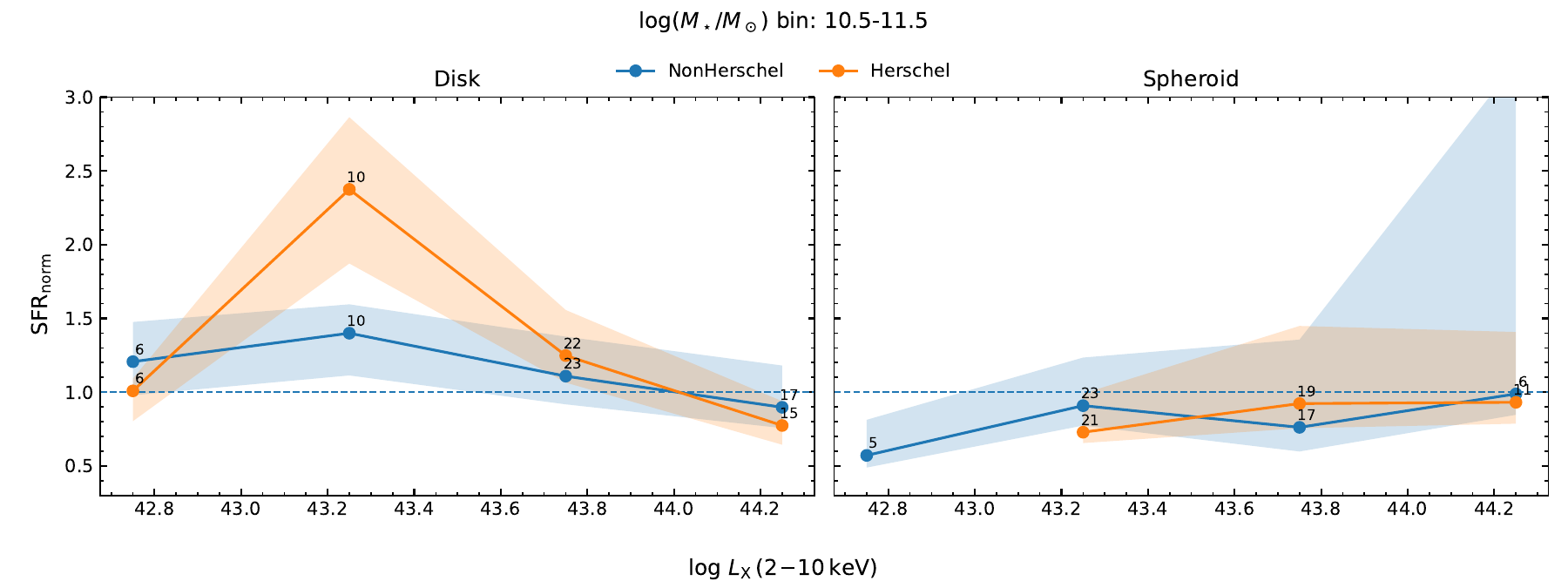}
\caption{Median SFR$_{norm}$ as a function of $L_{\mathrm{X}}$ in $M_\star$ range of $10.5 \leq \log(M_\star/M_\odot) < 11.5$, split by host-galaxy morphology.
Disk-dominated (left panel) and spheroid-dominated (right panel) systems are shown. Shaded regions indicatE the 16th–84th percentile ranges. Numbers next to the points give the number of AGN in each bin. The horizontal dashed line denotes SFR$_{\mathrm{norm}} = 1$.}
\label{fig:sfrnorm_lx_herschel_morph}
\end{figure*}

Here, we present the results of examining the impact of FIR photometry on the relation between $\mathrm{SFR}_{\rm norm}$ and $L_X$. Motivated by the findings of \citet{Mountrichas2025a}, we repeat the SFR$_{\mathrm{norm}}$ analysis separately for Herschel-detected and non-detected systems, considering both the full X-ray AGN sample and the morphology-split subsamples.

Following the methodology described in Sect.~\ref{sec:sfrnorm}, we compute SFR$_{\mathrm{norm}}$ for X-ray AGN using control samples of non-AGN galaxies matched in M$_\star$ and redshift. The analysis is performed independently for Herschel-detected and non-detected galaxies, while applying identical selections, M$_\star$ completeness weights, and quiescent-galaxy exclusion criteria in both cases.

Figure~\ref{fig:sfrnorm_lx_herschel} shows the SFR$_{\mathrm{norm}}$--$L_X$ relation for Herschel-detected and non-detected AGN host galaxies in bins of $M_\star$. Within the statistical uncertainties, we find no significant differences between the two subsamples across the explored $M_\star$ and $L_X$ ranges. The median SFR$_{\mathrm{norm}}$ values and their uncertainties are consistent in all $M_\star$ bins, and no systematic offsets or slope differences are observed.

We further verify this result by repeating the analysis separately for disk- and spheroid-dominated host galaxies. As shown in Fig.~\ref{fig:sfrnorm_lx_herschel_morph}, the morphology-dependent trends discussed in Sect.~\ref{sec:results} are recovered in both Herschel-detected and non-detected samples, with no statistically significant differences in either the normalisation or the shape of the SFR$_{\mathrm{norm}}$--$L_X$ relation.

These results indicate that, for the DESI--eRASS1 dataset and the analysis framework adopted in this work, FIR detectability does not play a dominant role in shaping the observed SFR$_{\mathrm{norm}}$--$L_X$ trends. This contrasts with the behaviour reported in \citet{Mountrichas2025a} and can be attributed to a combination of factors, including the explicit exclusion of quiescent galaxies in the present analysis and the higher typical AGN luminosities probed by the eRASS1 survey (see Sect.~\ref{sec:herschel}). We therefore conclude that splitting the sample by Herschel detection is not required for the main results presented in this paper.

\section{Impact of omitting NIR photometry}
\label{app:nir_impact}

In this Appendix, we investigate the effect of missing NIR photometry in the VAC catalogue on the three key parameters used throughout this work: $M_\star$, SFR, and $frac_{AGN}$. For this purpose, we perform a controlled comparison by running \textsc{CIGALE} twice on the same sample of sources, once including the available NIR photometry and once excluding it, and then directly comparing the derived parameter estimates. We also explain how these results are used to correct for the absence of NIR data where needed.

The NIR photometry used in this analysis consists primarily of the $H$ and $K$ bands from the available NIR catalogues (VIKING, VHS), which were cross-matched to the DESI sources using a matching radius of 1\arcsec. The comparison is based on the same set of galaxies in both runs, so that the only difference in the SED fitting is the inclusion or exclusion of the NIR bands. More than 90\% of the sources included in this comparison have both $H$ and $K$ photometry.

\subsection{Impact of omitting NIR photometry on $M_\star$ estimates}
\label{app:nir_impact_mstar}

\begin{figure}[t]
\centering
\includegraphics[width=0.49\textwidth]{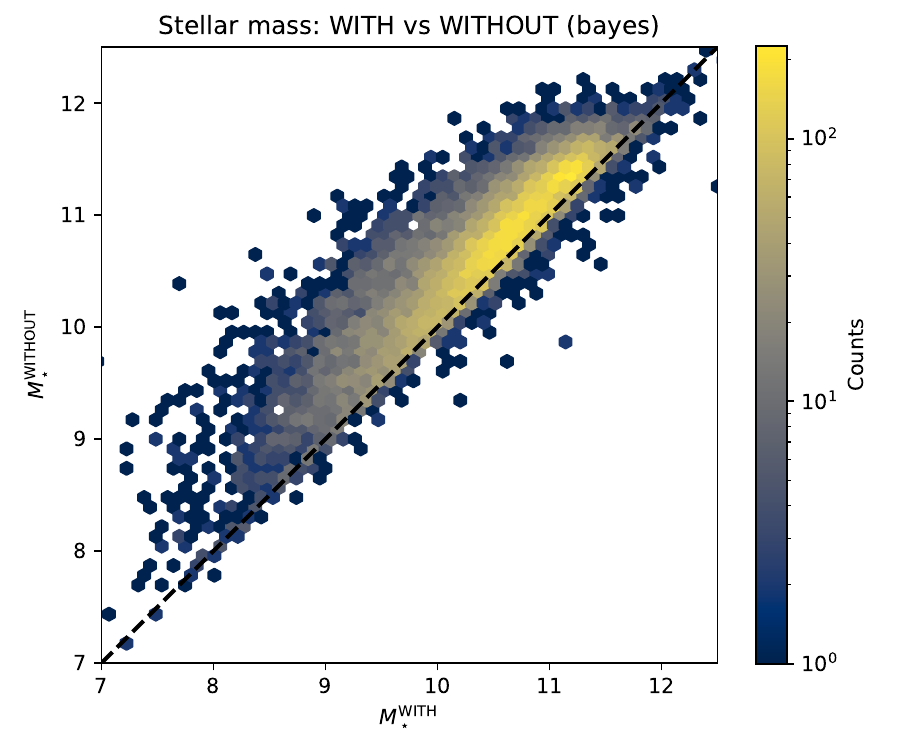}

\caption{Comparison between $M_\star$ inferred with NIR photometry (\emph{WITH NIR}) and without NIR photometry (\emph{WITHOUT NIR}). The black dashed line marks the one-to-one relation.}
\label{fig:nir_mstar}
\end{figure}

We first quantify how the absence of near-infrared (NIR) photometry affects the $M_\star$ estimates returned by \textsc{CIGALE}. Figure~\ref{fig:nir_mstar} compares the Bayesian $M_\star$ values obtained from fits including NIR constraints (hereafter \emph{WITH NIR}) with those obtained when the NIR bands are removed (\emph{WITHOUT NIR}). Although the two estimates remain tightly correlated, most sources lie above the one-to-one relation, showing that M$_\star$ inferred \emph{without} NIR photometry are systematically higher than those inferred \emph{with} NIR photometry. The median offset is $\Delta \log M_\star \simeq 0.24$ dex, with a scatter of $\simeq 0.24$ dex, where the latter is quantified using the normalized median absolute deviation (NMAD).

\begin{figure}[t]
\centering
\includegraphics[width=\columnwidth]{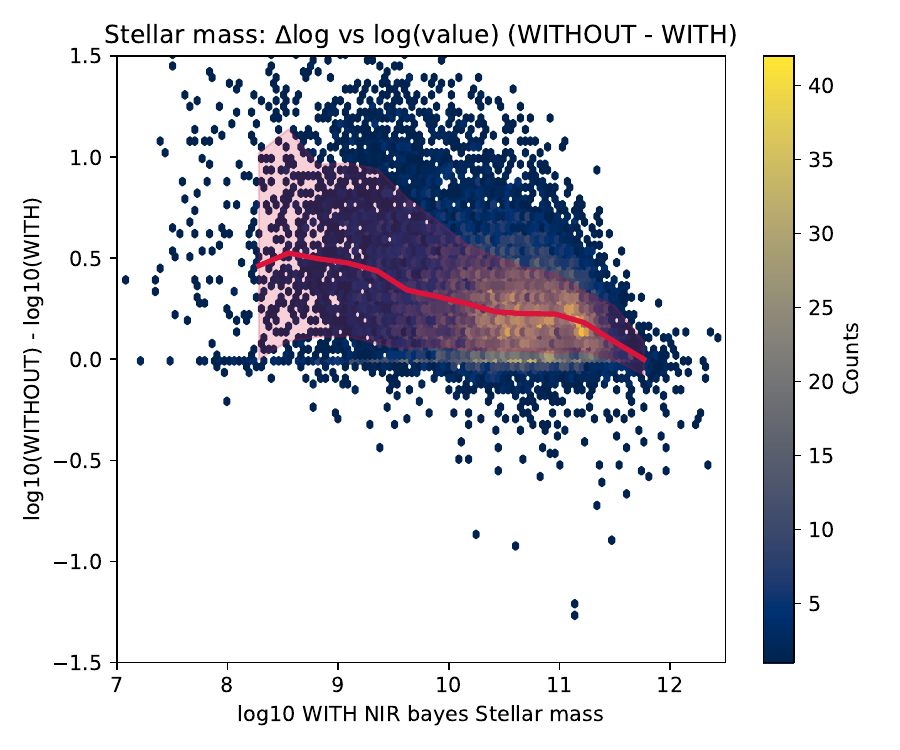}
\caption{Dependence of the $M_\star$ offset on $M_\star$. The solid line shows the running median of the relation. The offset is not constant, and the largest positive deviations preferentially occur at lower $M_\star^{\rm WITH}$, indicating that omitting NIR photometry affects the inferred stellar mass in a mass-dependent manner.}
\label{fig:nir_mstar_delta}
\end{figure}

Figure~\ref{fig:nir_mstar_delta} shows $\Delta \log M_\star$ as a function of $\log M_\star^{\rm WITH}$. The offset is not constant, but decreases with increasing $M_\star$, with the largest positive $\Delta \log M_\star$ values occurring preferentially at lower $M_\star^{\rm WITH}$ (red, solid line). This behaviour suggests that, for part of the sample, the absence of NIR data increases the degeneracy in the inferred stellar mass-to-light ratio, leading to systematically overestimated M$_\star$.

\begin{figure}[t]
\centering
\includegraphics[width=\columnwidth]{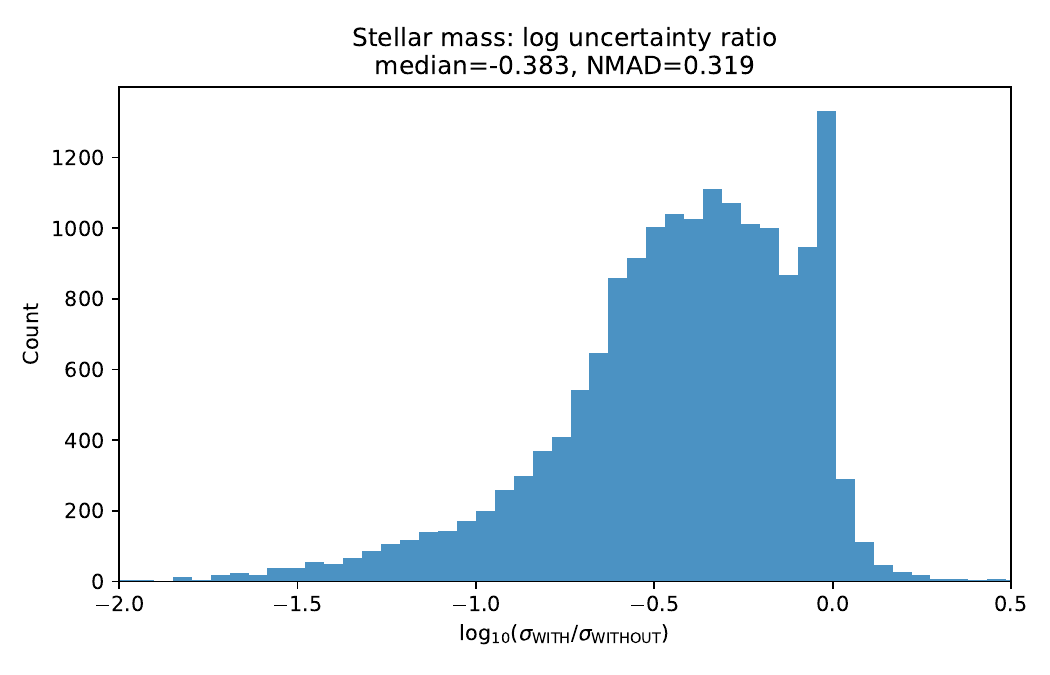}
\caption{
Comparison of Bayesian $M_\star$ uncertainties between the two fits.
The histogram shows $\log_{10}(\sigma_{\rm WITH}/\sigma_{\rm WITHOUT})$, where $\sigma$ is the Bayesian uncertainty on $M_\star$.
Negative values indicate larger uncertainties in the \emph{WITHOUT NIR} run.
}
\label{fig:nir_mstar_err}
\end{figure}

Finally, Fig.~\ref{fig:nir_mstar_err} compares the formal Bayesian uncertainties on $M_\star$ in the two runs through the distribution of $\log_{10}(\sigma_{\rm WITH}/\sigma_{\rm WITHOUT})$. The distribution is shifted toward negative values, with a median of $\simeq -0.38$ dex (NMAD $\simeq 0.32$ dex), indicating that the typical uncertainty on $M_\star$ is smaller when NIR photometry is included, by a factor of about 2.4. This highlights the additional constraining power of the NIR bands, which help break degeneracies in the stellar mass-to-light ratio and therefore yield more tightly constrained M$_\star$ estimates.

\subsection{Impact of omitting NIR photometry on SFR estimates}
\label{app:nir_impact_sfr}

\begin{figure}[t]
\centering
\includegraphics[width=0.49\textwidth]{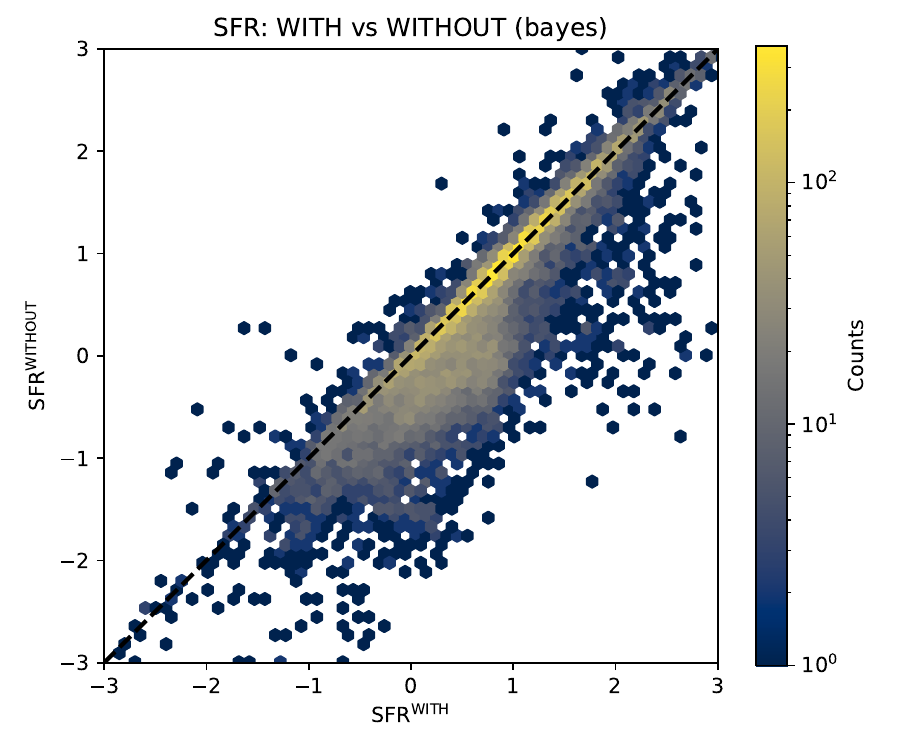}
\caption{Comparison between ${\rm SFR}$ inferred with and without NIR photometry.The solid line marks the one-to-one relation.}
\label{fig:nir_sfr}
\end{figure}

We next examine the effect of removing NIR photometry on the SFR inferred by \textsc{CIGALE}. Figure~\ref{fig:nir_sfr} compares the Bayesian SFR estimates obtained from fits including NIR data (\emph{WITH NIR}) with those obtained when the NIR bands are excluded (\emph{WITHOUT NIR}), on logarithmic axes. The two estimates remain strongly correlated and broadly follow the one-to-one relation over the full dynamic range. Noticeable deviations appear mainly at the lowest SFR values, where the scatter becomes larger, while no significant systematic offset is seen for the bulk of the sample. The median offset of $\Delta\log{\rm SFR} \simeq -0.04$ dex and a scatter (NMAD) of $\simeq 0.12$ dex, indicating that the absence of NIR data has only a minor effect on the overall SFR normalization.

\begin{figure}[t]
\centering
\includegraphics[width=\columnwidth]{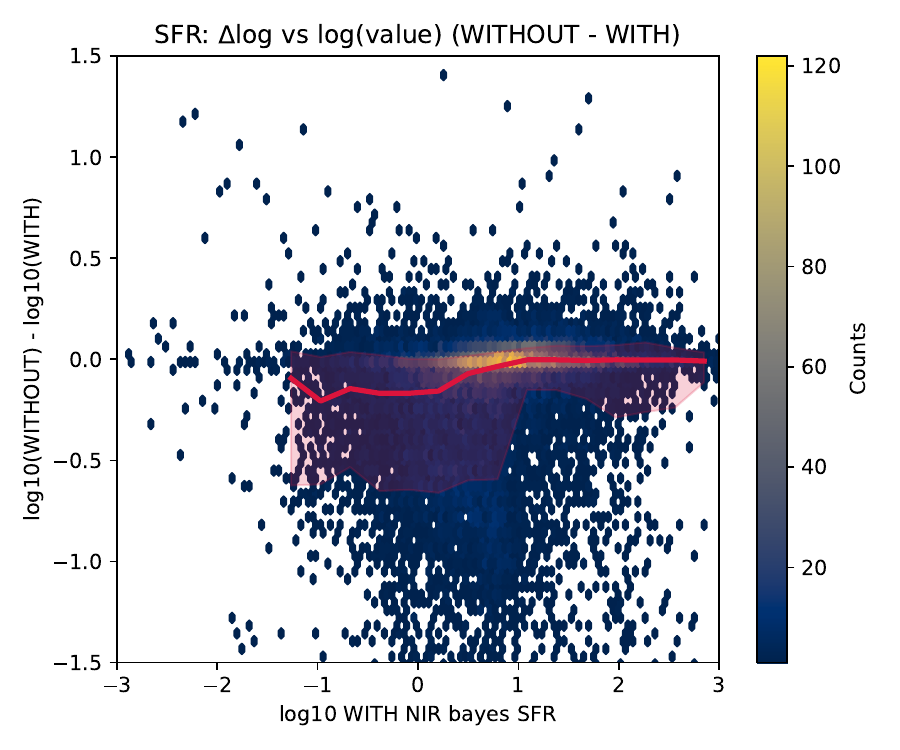}
\caption{
Difference in SFR estimates as a function of ${\rm SFR}^{\rm WITH}$. The solid line shows the running median of the relation.
Only mild deviations are observed at the lowest SFR values, while no systematic trend is present over the bulk of the sample.
}
\label{fig:nir_sfr_delta}
\end{figure}

Figure~\ref{fig:nir_sfr_delta} further shows $\Delta\log{\rm SFR}$ as a function of $\log{\rm SFR}^{\rm WITH}$. Over most of the SFR range, the offset remains consistent with zero, while at the lowest SFR values the scatter increases and the running trend shows a mild negative deviation. This behaviour suggests that any impact from omitting NIR data is confined mainly to the low-SFR regime and does not introduce a global bias in the inferred SFRs.

\begin{figure}[t]
\centering
\includegraphics[width=\columnwidth]{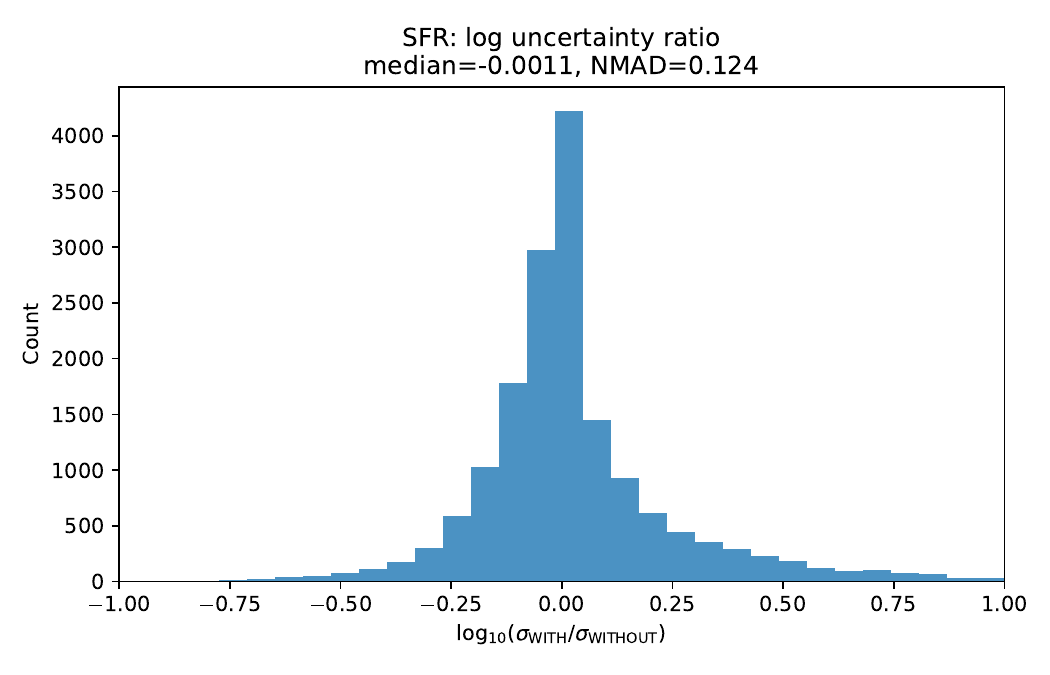}
\caption{
Comparison of Bayesian SFR uncertainties between the two fits.
The histogram shows $\log_{10}(\sigma_{\rm WITH}/\sigma_{\rm WITHOUT})$, indicating nearly identical uncertainty distributions in the two cases.
}
\label{fig:nir_sfr_err}
\end{figure}

Finally, Fig.~\ref{fig:nir_sfr_err} compares the formal Bayesian uncertainties on the SFR between the two runs through the distribution of $\log_{10}(\sigma_{\rm WITH}/\sigma_{\rm WITHOUT})$. This distribution is narrowly centered on zero, with a median of $\simeq -0.001$ dex and an NMAD of $\simeq 0.12$ dex, indicating that the inclusion of NIR photometry does not significantly modify the uncertainty budget of the SFR estimates.

Overall, we conclude that the SFR estimates used in this work are robust to the absence of NIR photometry, and no correction is required when comparing the \emph{WITH NIR} and \emph{WITHOUT NIR} datasets.

\subsection{Impact of omitting NIR photometry on $frac_{AGN}$ estimates}
\label{app:nir_impact_fagn}

\begin{figure}[t]
\centering
\includegraphics[width=0.49\textwidth]{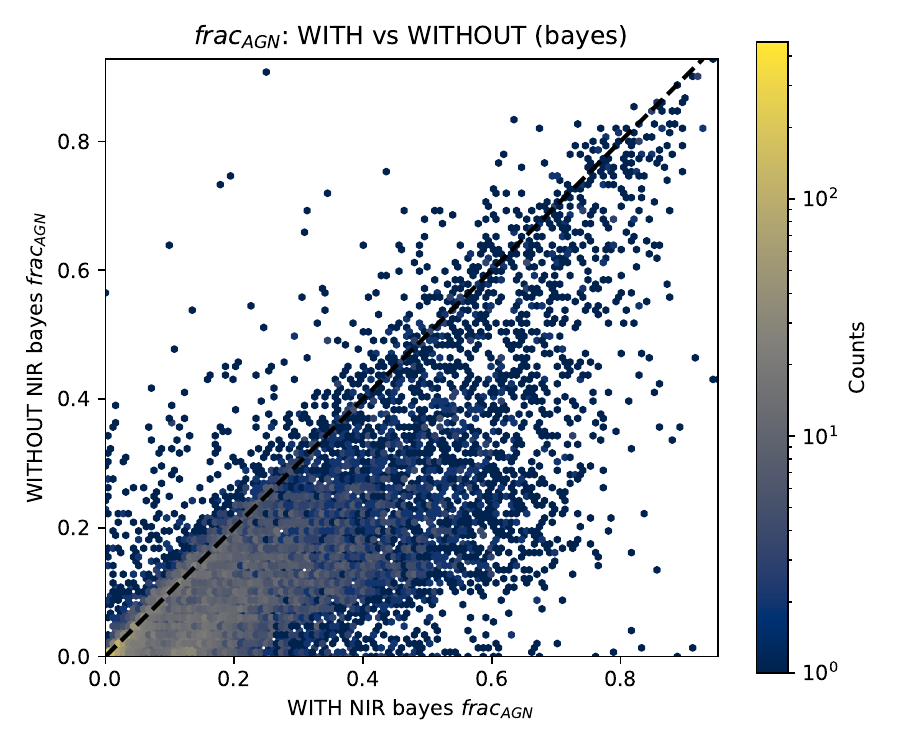}
\caption{
Comparison between $frac_{AGN}$ inferred with and without NIR photometry. The solid line marks the one-to-one relation.}
\label{fig:nir_fagn}
\end{figure}

Finally, we assess the impact of removing NIR photometry on the $frac_{AGN}$ parameter inferred by \textsc{CIGALE}. Figure~\ref{fig:nir_fagn} compares the Bayesian $frac_{AGN}$ values obtained from fits including NIR data (\emph{WITH NIR}) with those obtained when the NIR bands are excluded (\emph{WITHOUT NIR}). Although a clear correlation is present, most sources lie below the one-to-one relation, indicating that the AGN fraction inferred without NIR photometry is systematically lower than that inferred when NIR constraints are included. This systematic difference has a median $\Delta frac_{AGN} \simeq -0.07$ and a scatter (NMAD) of $\simeq 0.11$, showing that the absence of NIR data leads to a non-negligible underestimation of the AGN contribution.

\begin{figure}[t]
\centering
\includegraphics[width=\columnwidth]{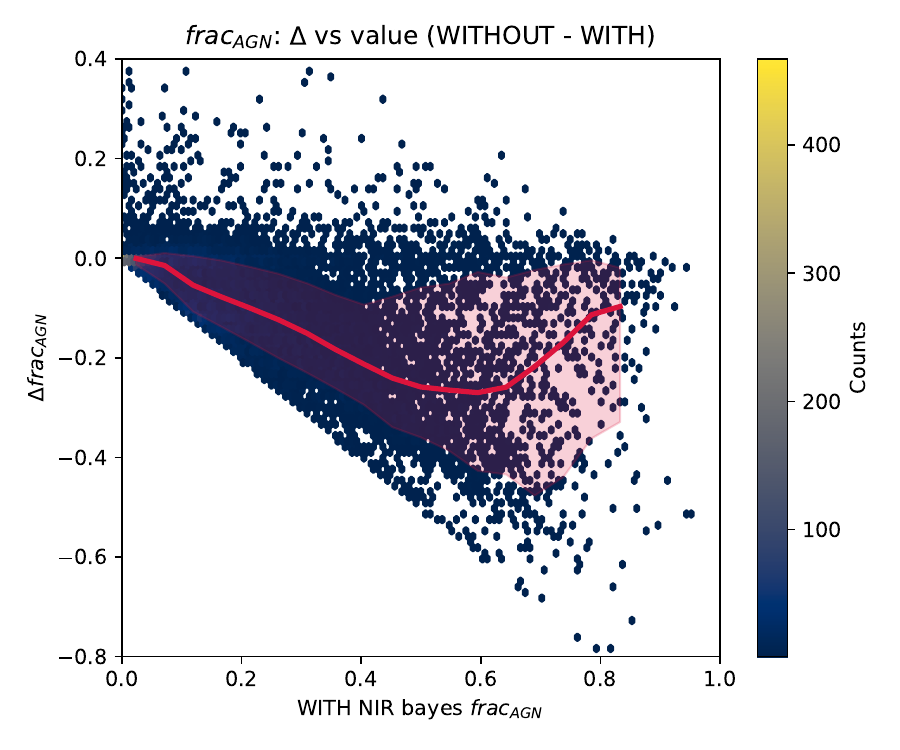}
\caption{
Difference in AGN fraction as a function of ${\rm f_{AGN}}^{\rm WITH}$.
The solid line shows the running median of the relation. The offset depends on the AGN fraction itself, with the strongest deviations occurring at intermediate values, reflecting increased AGN–host degeneracy in the absence of NIR photometry.
}
\label{fig:nir_fagn_delta}
\end{figure}

Figure~\ref{fig:nir_fagn_delta} further shows $\Delta frac_{AGN}$ as a function of $frac_{AGN}^{\rm WITH}$. The offset depends strongly on the AGN fraction itself: the median trend becomes increasingly negative from low to intermediate $frac_{AGN}^{\rm WITH}$, reaches its largest deviation at intermediate-to-high values, and then partially flattens toward the highest AGN fractions. This behaviour suggests that the absence of NIR constraints enhances the degeneracy between stellar and AGN emission in the SED fitting, particularly for sources in which the AGN contributes significantly but does not fully dominate the observed SED. In this regime, the NIR bands help constrain the relative contributions of the stellar continuum and the AGN component, thereby reducing ambiguities in the inferred $frac_{AGN}$.

\begin{figure}[t]
\centering
\includegraphics[width=\columnwidth]{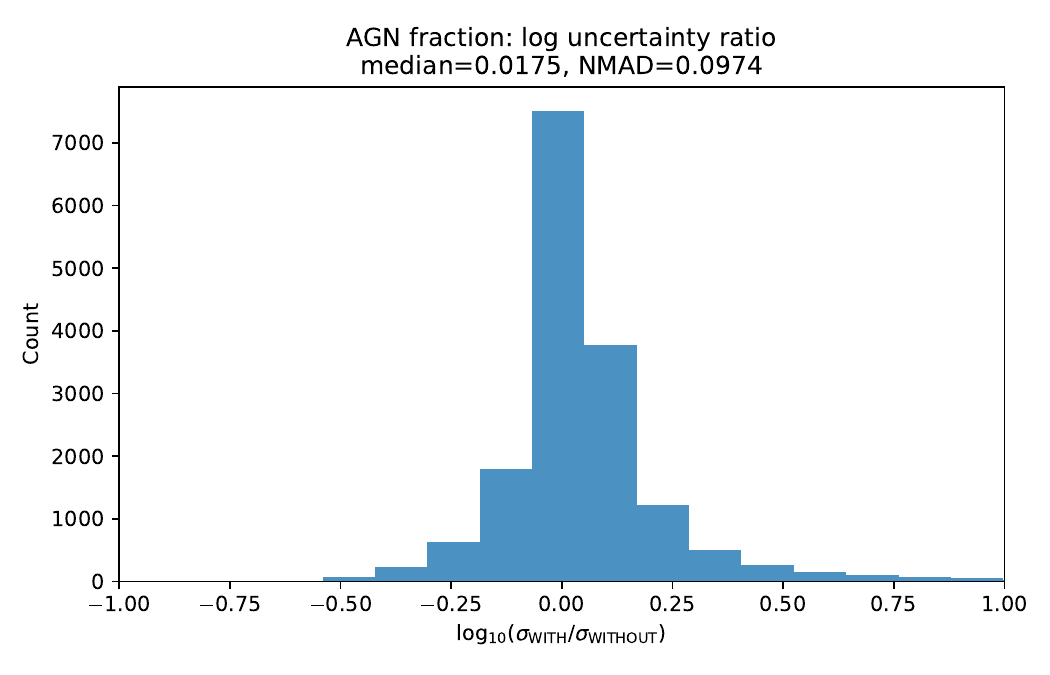}
\caption{
Comparison of Bayesian AGN fraction uncertainties between the two fits.
The histogram shows $\log_{10}(\sigma_{\rm WITH}/\sigma_{\rm WITHOUT})$, indicating similar uncertainty distributions despite systematic offsets in the inferred values.
}
\label{fig:nir_fagn_err}
\end{figure}

Finally, Fig.~\ref{fig:nir_fagn_err} compares the formal Bayesian uncertainties on $frac_{AGN}$ between the two runs through the distribution of $\log_{10}(\sigma_{\rm WITH}/\sigma_{\rm WITHOUT})$. This distribution is narrowly centered close to zero, with a median of $\simeq 0.02$ dex and an NMAD of $\simeq 0.10$ dex, indicating that the systematic differences between the two estimates are not driven by a substantial change in the formal uncertainty, but instead reflect shifts in the inferred posterior distributions. Overall, similarly to the case of $M_\star$, the $frac_{AGN}$ estimates are sensitive to the inclusion of NIR photometry.

\subsection{Correction scheme for the absence of NIR photometry}
\label{app:nir_corrections}

Based on the diagnostics presented above, we define empirical corrections that map $M_\star$ and $frac_{AGN}$ estimates obtained without NIR photometry onto the reference scale defined by the fits including NIR data.
These corrections are applied to ensure a homogeneous parameter scale across the full sample, while no correction is applied to the SFR. 

We note that the availability of NIR photometry in the VAC catalogue is driven by observational coverage rather than by intrinsic source properties. We verified that the availability of NIR photometry is not associated with systematic differences in basic observational or physical properties.
In particular, we find no evidence that sources lacking NIR photometry are preferentially fainter or occupy a distinct redshift range compared to sources with NIR coverage.
Similarly, their distributions in $M_\star$, SFR, and $frac_{AGN}$ broadly overlap with those of the NIR subsample.
These checks support the interpretation that the absence of NIR data reflects observational coverage rather than intrinsic source differences. The empirical corrections derived from the NIR subsample therefore capture the effect of missing NIR constraints, rather than differences between distinct source populations, and can be safely applied to the full sample.

For the $M_\star$, we work in logarithmic space and define the offset
$\Delta\log M_\star \equiv \log M_\star^{\rm WITH} - \log M_\star^{\rm WITHOUT}$.
We derive a binned correction by computing the median $\Delta\log M_\star$ as a function of $\log M_\star^{\rm WITHOUT}$, requiring at least 200 sources per bin.
This correction is then applied to the WITHOUT--NIR $M_\star$, yielding corrected values that are intended to reproduce the WITH--NIR estimates.

\begin{figure}[t]
\centering
\includegraphics[width=\columnwidth]{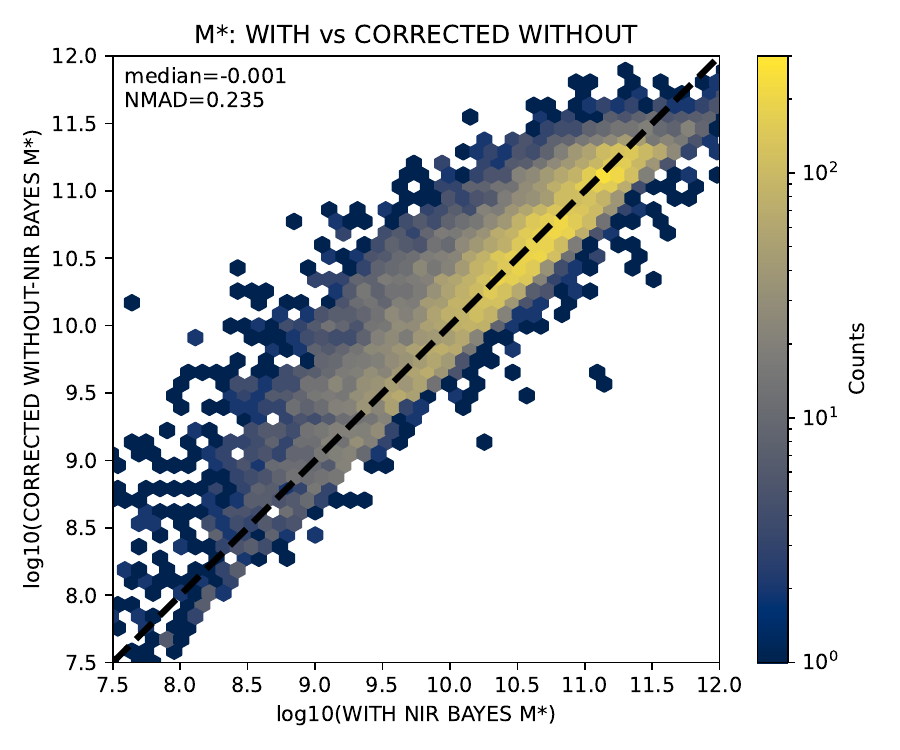}
\caption{
Comparison between M$_\star$ inferred with NIR photometry and those inferred without NIR photometry after applying the empirical correction.
The dashed line marks the one-to-one relation.
}
\label{fig:nir_mstar_corr}
\end{figure}

\begin{figure}[t]
\centering
\includegraphics[width=\columnwidth]{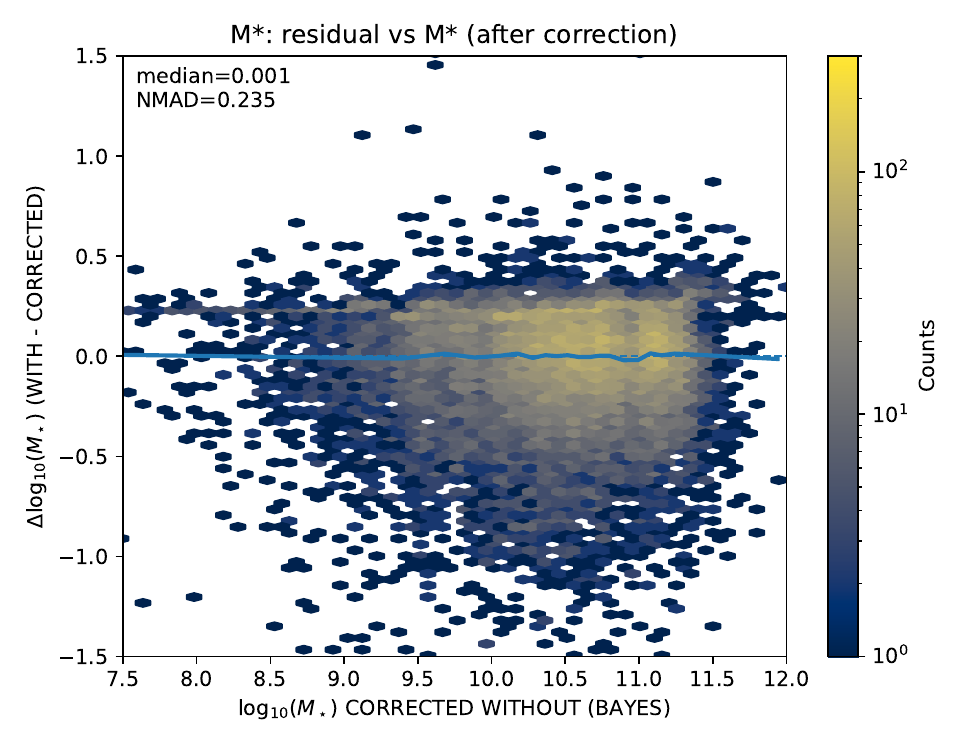}
\caption{
Residuals between $M_\star^{\rm WITH}$ and corrected $M_\star^{\rm WITHOUT}$ as a function of M$_\star$. The solid line shows the running median of the relation.
No systematic trend remains after applying the correction.
}
\label{fig:nir_mstar_corr_res}
\end{figure}

Figure~\ref{fig:nir_mstar_corr} shows the comparison between $M_\star^{\rm WITH}$ and the corrected $M_\star^{\rm WITHOUT}$.
The two estimates closely follow the one-to-one relation, with a median residual of $\simeq 0.0$ dex and an NMAD of $\simeq 0.24$ dex.
The residuals as a function of $M_\star$ (Fig.~\ref{fig:nir_mstar_corr_res}) show no remaining systematic trend, demonstrating that the mass-dependent bias introduced by the absence of NIR photometry is effectively removed.

A similar approach is adopted for the $frac_{AGN}$.
In this case, the correction is defined in linear space as
$\Delta frac_{AGN} \equiv frac_{AGN}^{\rm WITH} - frac_{AGN}^{\rm WITHOUT}$,
and is derived as a binned median function of $frac_{AGN}^{\rm WITHOUT}$, again requiring at least 200 sources per bin.
To avoid over-correction at the highest $frac_{AGN}$, where the offset naturally flattens, the correction is smoothly tapered at large $frac_{AGN}$ values.

\begin{figure}[t]
\centering
\includegraphics[width=\columnwidth]{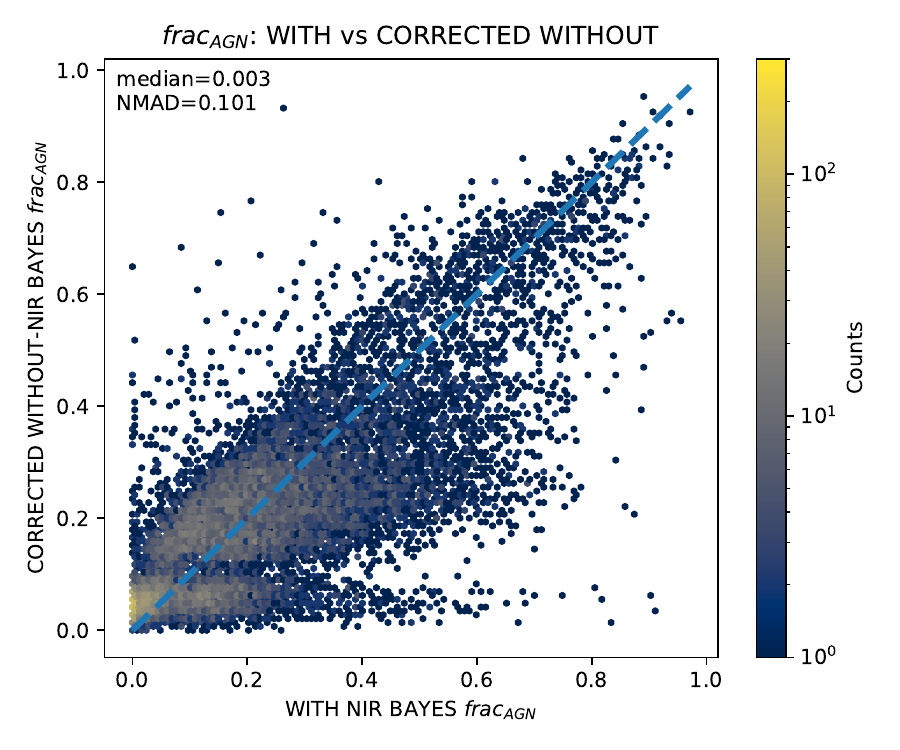}
\caption{
Comparison between $frac_{AGN}$ inferred with NIR photometry and those inferred without NIR photometry after applying the empirical correction.
The dashed line marks the one-to-one relation.
}
\label{fig:nir_fagn_corr}
\end{figure}

\begin{figure}[t]
\centering
\includegraphics[width=\columnwidth]{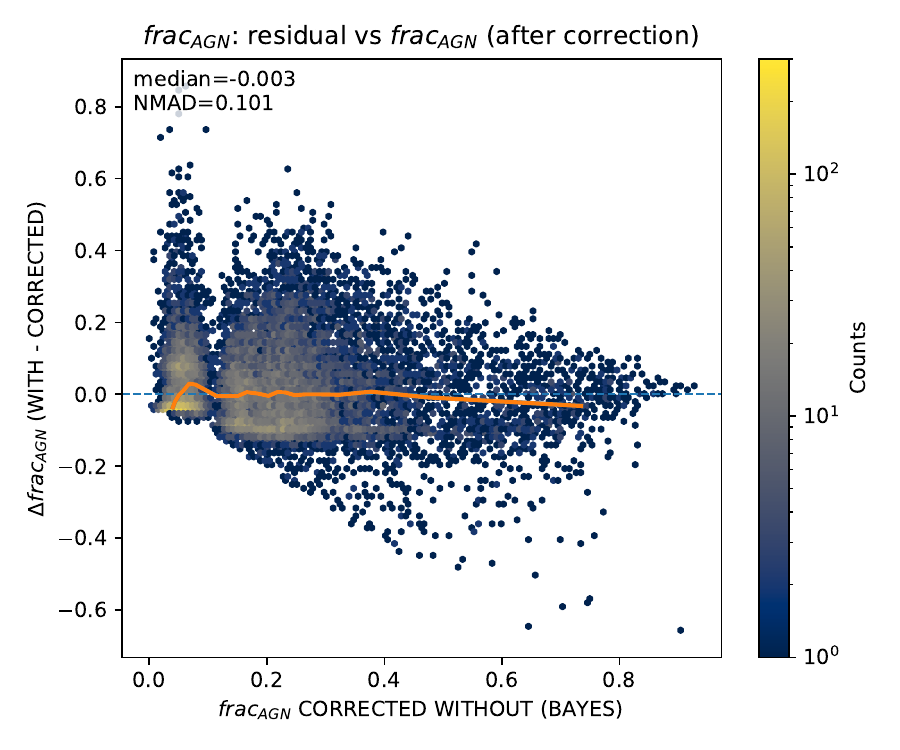}
\caption{
Residuals between $frac_{AGN}^{\rm WITH}$ and corrected $frac_{AGN}^{\rm WITHOUT}$ as a function of $frac_{AGN}$. The solid line shows the running median of the relation.
The absence of residual trends confirms the effectiveness of the correction.
}
\label{fig:nir_fagn_corr_res}
\end{figure}

Figure~\ref{fig:nir_fagn_corr} shows the comparison between $frac_{AGN}^{\rm WITH}$ and the corrected $frac_{AGN}^{\rm WITHOUT}$.
After correction, the two estimates are consistent, with a median offset consistent with zero and an NMAD of $\simeq 0.10$. The two partially separated density ridges visible after the correction arise from the non-linear mapping applied to the original $frac_{AGN}^{\rm WITHOUT}$ distribution. Because the empirical correction depends on $frac_{AGN}^{\rm WITHOUT}$ and flattens at high values, the transformation of the compressed WITHOUT--NIR distribution produces two apparent loci in the corrected plane. This effect reflects the functional form of the correction rather than the presence of distinct physical populations. The residuals as a function of $frac_{AGN}$ (Fig.~\ref{fig:nir_fagn_corr_res}) show no significant systematic trend, confirming that the value-dependent bias identified in the absence of NIR photometry is removed.

The corrected $M_\star$ and $frac_{AGN}$ are adopted throughout the analysis to ensure a homogeneous parameter scale across the full sample, independent of the availability of NIR photometry for individual sources.

\end{document}